# Grand Challenges for the Convergence of Computational and Citizen Science Research

February 2026

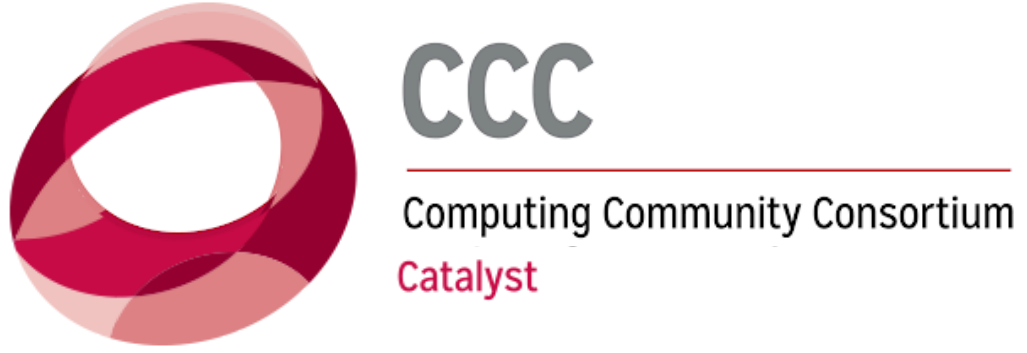

## Authors

### Lead Authors

Lucy Fortson (University of Minnesota)
Lea Shanley (International Computer Science Institute and GNIES, University of Wisconsin-Madison)
Tanya Berger-Wolf (The Ohio State University)
Kevin Crowston (Syracuse University)
Corey Jackson (University of Wisconsin-Madison)
Saiph Savage (Northeastern University)
Haley Griffin (Computing Research Association)

### Key Contributing Authors

Sara Beery, Massachusetts Institute of Technology
Muki Haklay, University College London
Aggelos Katsaggelos, Northwestern University
Harmanpreet Kaur, University of Minnesota - Twin Cities
Thai Le, Indiana University
Chris Lintott, University of Oxford
Pietro Michelucci, Human Computation Institute
Dara Norman, NOIRLab
Julia K Parrish, University of Washington
Jonathan Romano, Frameshifter
Michela Taufer, University of Tennessee-Knoxville
Laura Trouille, Adler Planetarium - Zooniverse

### Collaborators

All Workshop and Roundtable Participants provided substantial intellectual input - please see lists of participants in Appendix B & C.

### Acknowledgements

We would like to thank Michela Taufer (University of Tennessee-Knoxville) for serving as the CCC Council liaison on the workshop organizing team, and reviewing the final report. We would like to thank Elora Daniels (Computing Research Association) for figure design. In addition to the above contributing authors, we also would like to thank those participants who provided feedback on the initial draft of the report: Nataly Buslón (Pompeu Fabra University), Sola Kim (Arizona State University), Daniel Lee (Schmidt Sciences), Yiftach Nagar (Academic College of Tel Aviv-Yaffo), and Kat Schrier (Marist University).

## Suggested Citation

Fortson, L., Shanley, L., Berger-Wolf, T., Crowston, K., Jackson, C., Savage, S., Griffin, H. (2026) *Grand Challenges for the Convergence of Computational and Citizen Science Research*. Washington, D.C.: Computing Research Association (CRA). http://cra.org/ccc/wp-content/uploads/sites/2/2026/02/grand-challenges-for-the-convergence-of-computational-and-citizen-science-research. Available at SSRN https://ssrn.com/abstract=6730679

## About the Computing Community Consortium (CCC)

A programmatic committee of the Computing Research Association (CRA), CCC enables the pursuit of innovative, high-impact computing research that aligns with pressing national and global challenges. Of, by, and for the computing research community, CCC is a responsive, respected, and visionary organization that brings together thought leaders from industry, academia, and government to articulate and advance compelling research visions and communicate them to stakeholders, policymakers, the public, and the broad computing research community.

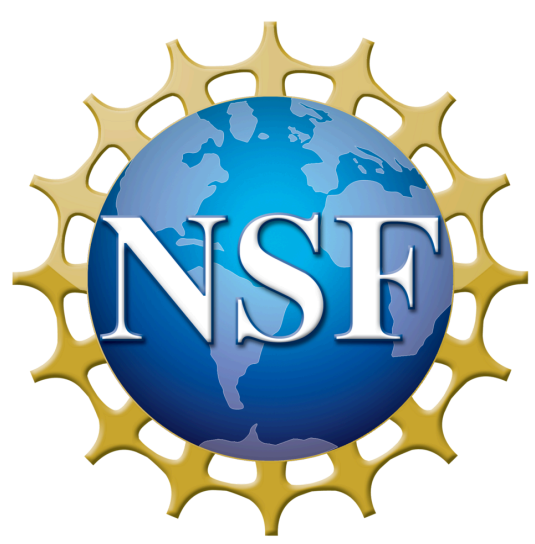

*This material is based upon work supported by the U.S. National Science Foundation (NSF) under Award No. 2300842. These awards support the Computing Community Consortium (CCC), a programmatic committee of the Computing Research Association (CRA).*

*Any opinions, findings, and conclusions or recommendations expressed in this material are those of the author(s) and do not necessarily reflect the views of the National Science Foundation.*

# TABLE OF CONTENTS

## EXECUTIVE SUMMARY

Over the past several decades, scientists and communities have increasingly worked together through citizen science[1] to advance scientific research and address real-world problems. This approach allows volunteers to collect data across wide geographic areas and over long periods of time, filling critical gaps that professional scientists and agencies often cannot cover alone. In addition, the citizen science approach has accelerated the processing of massive scientific datasets through volunteer labeling efforts and distributed volunteer computation. At the same time, revolutionary strides in computation, machine learning, and artificial intelligence (ML/AI) technologies, as well as technology platforms for large-scale deployment of digital tools and computation infrastructure, are significantly transforming the way we do science. Together, these developments create a powerful opportunity whereby research in citizen science can augment computational research and vice versa. **By combining the speed and scale of modern computational methods and technologies with the reach of public participation as a model of human-centered computing, we can accelerate discovery, inform decision-making, and make science more responsive to local and national needs.** The unique and evolving relationship between citizen science and AI is set to generate more scientific breakthroughs, lower barriers in making science more accessible to all, and democratize science in a way that provides benefits to society as a whole. Developing the machines of tomorrow, integrated with human talent at the core, will unlock future frontiers in science and citizen science currently beyond our imagination.

This report is an outcome of a Computing Community Consortium (CCC) visioning workshop on **Grand Challenges for the Convergence of Computational and Citizen Science Research** conducted on April 8-9, 2025, in Washington, D.C. as well as through several precursor virtual input-gathering sessions. These events brought together experts across relevant disciplines to develop a research agenda that brings to fruition the above vision on how humans and machines may team up to solve some of the world's most pressing scientific problems.

The convergence of citizen science and computing research aligns with federal directives and national interests. Convergence supports mandates in AI governance, national competitiveness, transparency, and efficient public service delivery, including "Accelerating Federal Use of AI through Innovation, Governance, and Public Trust" (M-25-21) and "Driving

---

[1] According to the Crowdsourcing and Citizen Science Act of 2016 (15 U.S. Code § 3724), citizen science is defined as an open collaboration in which individuals or organizations voluntarily participate in the scientific process — such as formulating research questions, collecting or analyzing data, developing technologies, or interpreting results — while crowdsourcing refers to obtaining services, ideas, or content through voluntary contributions from a group, especially via online platforms. In computer science, crowdsourcing can be associated with paying multiple, independent contributors to perform micro-tasks.

Efficient Acquisition of Artificial Intelligence in Government" (M-25-22). We also referenced National Academies of Sciences, Engineering, and Math (NASEM) reports, such as the "Human-AI Teaming: State-of-the-Art and Research Needs (2022)" [hereafter NASEM 2022].

Citizen science delivers measurable economic and national value. Public participation in scientific research generates millions of dollars in volunteer labor value, extends government agency capacity, and directly supports federal priorities in areas such as disaster management, public health, water, energy, workforce development, and many more.

At the same time, 21st-century scientific infrastructure requirements for citizen science (from hardware and cyberinfrastructure to data and computational frameworks) mirror those for computational science more generally. The distributed, collaborative, long-term, and contextual nature of citizen science makes it a demanding real-world use case for a novel robust research infrastructure that accounts for security, privacy, resource adaptability, and transparency.

Below, we outline the key findings, future research directions, and recommendations that emerged from the April 2025 CCC Grand Challenges for the Convergence of Computational and Citizen Science Research Workshop.

## Key Findings

The convergence of computational and citizen science research represents a generational opportunity to reimagine how we conduct research, involve the public, and deliver scientific value to society. This CCC workshop surfaced a rich set of ideas for near-term investment and long-term strategic alignment. Rather than treating citizen science and computational science as parallel tracks, the workshop reframed them as interdependent pillars of 21st-century science, each strengthening the other when thoughtfully integrated. Key aspects of this interdependency include:

1. **Mutual reinforcement of AI/ML and citizen science:** A positive feedback loop exists between citizen science and AI/ML in which each can augment and enhance the other, enabling entirely new use cases. Using crowdsourcing and human-in-the-loop methodologies, as well as feedback to models and systems, citizen science benefits AI/ML models by providing trusted, contextualized data across multiple scales. AI/ML benefits citizen science by providing optimized task routing, real-time coaching, and feedback, for example, through the judicious deployment of Large Language Models.

2. **Human–computer teaming is essential:** Effective teaming requires balancing machine efficiency and full parameter exploration with human insight and interests, especially in multi-agent, multi-modal, and multi-setting contexts. Humans can catch and explain anomalies, biases, and contextual errors that AI alone misses, interpreting and deciding what is relevant for scientific purposes.

3. **Feedback and interactivity are critical for engagement and data quality:** Personalized, real-time, reciprocal feedback (e.g., via AI tutors, multilingual support, iterative design) is needed for volunteer retention, accuracy, and learning.

4. **Trust is fragile but foundational:** Trust in computational systems hinges on transparency, participatory governance, explainable AI, and clear metrics of mutual respect and accountability between system developers and project participants.

At the same time, we must address key challenges:

5. **Infrastructure as a bottleneck and opportunity:** The needs of citizen science exceed the capabilities of current cyberinfrastructure, networks, and data systems more than most scientific domains, exposing the need for next-generation infrastructure (e.g., edge devices, lightweight AI models, decentralized and FAIR data systems, long-term sustainability models). Novel infrastructure could allow for dynamic scaling that is currently not possible, could accelerate science, and provide additional context and new data.

6. **Security, privacy, and adversarial threats are increasing risks:** Citizen science platforms face growing risks from synthetic media, data poisoning, and cyberattacks. Privacy-preserving and sovereignty-respecting governance models are underdeveloped.

7. **Momentum is real but fragmented:** Many successful projects that combine citizen science and computational research exist (e.g., iNaturalist, Mesonet, Stall Catchers, and Zooniverse.org), but lack shared standards, sustainable resourcing (especially for platform support), and coordinated governance – limiting scalability. Further, there is a knowledge-sharing gap between computational researchers and researchers engaged in integrating computational and citizen science, even though they are often investigating similar problems such as in the area of human-centered computing.

## Research Priorities to Meet National Needs

After exploring the needs for convergence of computational and citizen science, workshop participants discussed what was missing for convergence and then defined five *research drivers* that set the strategic priorities for advancing the convergence of citizen science and computation (see figure below).

In summary, citizen science can be transformed when **humans and machines work as complementary partners** (Driver 1), but this requires strong **feedback systems** that keep volunteers engaged and learning (Driver 2). For such collaboration to succeed, it must be built on **trust** — ensuring governance, transparency, and accountability in AI-enabled systems (Driver 3). At the same time, systems must remain **open yet secure**, balancing accessibility with protections for privacy, security, and data sovereignty (Driver 4). Finally, all of this depends on building and maintaining **sustainable infrastructure and long-term community support** — from adaptable project platforms, edge devices, and lightweight AI models to decentralized data systems, connectivity, and long-term community support (Driver 5).

Taken together, the drivers define a **comprehensive research agenda for the convergence of citizen science with advanced computing and AI**. Research carried out through the five drivers will lead to a socio-technical ecosystem where citizen participation and computational systems reinforce each other. Success will depend not only on technical innovation, but also on responsible governance, sustained infrastructure, and continuous human engagement.

# A 10-Year Community Roadmap for Convergence of Computational and Citizen Science

## Societal Benefits

- Provides human-centered, adaptive infrastructure for large-scale, multi-modal data collection and analysis
- Accelerates scientific discovery and innovation, boosting national competitiveness
- Harnesses public insight and data to tackle complex challenges in health, environment, and disaster response
- Cultivates a technically literate population

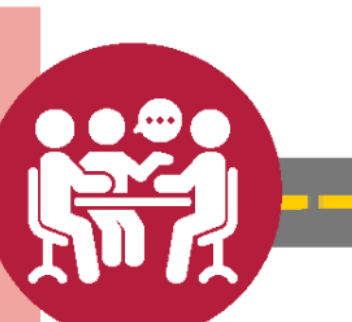

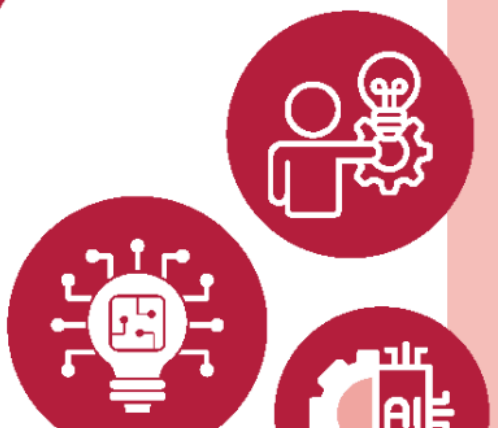

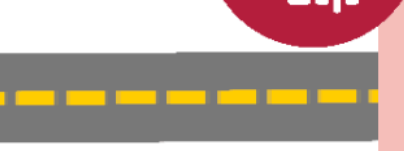

## What's Missing for Convergence?

- Human-computer teaming for public participation
- Explainable AI, particularly for non-experts
- Real-time engagement-optimized feedback systems
- Scalable, privacy-preserving data sharing and governance
- Sustainable civic science infrastructure
- Participatory AI governance through co-creation
- Trust metrics and long-term engagement research
- Cross-disciplinary training initiatives

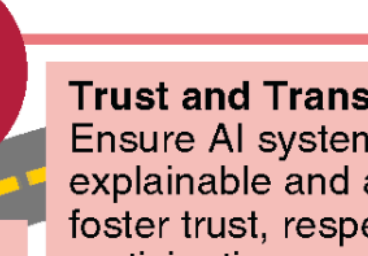

## Research Priorities to Meet Local and National Needs

**Human-AI Teaming:** Design multi-agent collaborative systems to optimize task sharing, interpretability, and adaptability across contexts.

**Trust and Transparency:** Ensure AI systems are explainable and accountable to foster trust, respect, and ethical participation.

**Real-Time Feedback:** Build interactive platforms with just-in-time, personalized feedback to boost learning, data quality, and engagement.

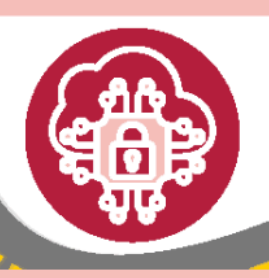

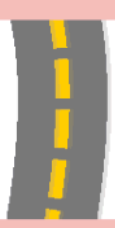

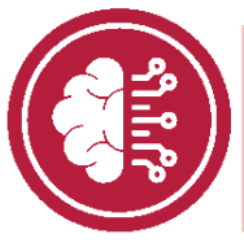

**Privacy and Security:** Protect citizen data while enabling open, resilient, and secure participatory science at scale.

**Sustainable Infrastructure:** Develop adaptable systems across cyber, data, human, and physical layers to support long-term, scalable citizen science.

## Recommendations

**National Infrastructure for Convergence:** developing and sustaining platforms, governance systems, and physical/cyber architecture required to support scalable, trustworthy, and nationwide convergence efforts.

**Core Research for Convergence:** focuses on foundational scientific and socio-technical investigations including developing, piloting and implementing new models, metrics, and frameworks for human-AI interaction, trust, and accountability.

**Training and Capacity Building:** developing the human capital—the skills, knowledge, and organizational structures—needed to create, manage, and participate in convergence projects across all sectors.

## Recommendations

Clear leadership across federal, state, and tribal governments with targeted investments are essential for the United States and its communities — from rural to urban — to fully capitalize on these opportunities and address the challenges. Our recommendations are grouped into three major themes.

### National Infrastructure for Convergence

This theme focuses on the sustained platforms, governance systems, and physical/cyber architecture required to support scalable, trustworthy, and nationwide convergence efforts.

1. **Encourage and incentivize cross-agency collaboration for convergence projects:** Break down silos by integrating citizen science and crowdsourcing with government agency AI, cloud, and technology strategies.

2. **Create permanent federal funding streams for convergence infrastructure:** Ensure stable funding for long-term platform and infrastructure maintenance and evolution.

3. **Build interoperable and AI-ready data infrastructure for participatory sciences:** Define national standards (FAIR data, APIs, metadata) for data exchange across platforms and projects.

4. **Develop scalable provenance frameworks for participatory AI systems:** Fund research on lightweight, privacy-aware provenance capture that spans human actions, AI model evolution, and distributed execution, enabling reproducibility, auditing, and long-term reuse at the national scale (Supported by findings 5, 6).

5. **Invest in privacy-preserving and sovereignty-respecting data frameworks:** Implement and evaluate mechanisms (e.g., federated learning, data trusts, open standards) that are core to participatory science infrastructure, and formalize the necessary governance and legal structures for secure data use.

6. **Establish national guidance for explainable, transparent, and trustworthy AI in citizen science:** Develop research-backed guidelines for interpretable outputs, community consent, and auditability.

7. **Develop next-generation participatory AI governance:** Co-create frameworks that give participatory science communities real agency over AI deployment decisions, data use, and model evolution.

8. **Launch a National Citizen Science & AI Convergence Hub:** Establish a central, virtual hub to share tools, standards, best practices, case studies, and training materials, thus reducing duplication and accelerating adoption.

### Core Research for Convergence

This theme covers the foundational scientific and socio-technical investigations required to advance the field, focusing on developing new models, metrics, and frameworks for human-AI interaction, trust, and accountability.

1. **Encourage and incentivize cross-disciplinary and cross-sectoral collaboration for convergence projects:** Implement strategies to break down silos and incentivize research and knowledge sharing across academic, industry, and community sectors.

2. **Develop Human-AI teaming frameworks for public participation:** Prototype and study new models for complementary collaboration between AI and citizen scientists, particularly incorporating large language models.

3. **Develop explainable AI for non-expert users:** Research novel user interface and data visualization techniques to make AI decision-making transparent and interpretable to the general public, enabling non-experts to better understand how AI is used. In turn, this work will also benefit ML and domain experts.

4. **Design real-time feedback systems for citizen science:** Research systems that use LLMs, AR/VR, edge computing, and mobile-first design for digital feedback loops to improve data quality, engagement, and learning outcomes.

5. **Institutionalize evaluation and trust metrics:** Fund research on trust diagnostics, engagement dynamics, and societal benefit indicators to guide iterative improvement and accountability.

6. **Advance participatory AI governance models:** Develop mechanisms for community consent, opt-out, and the co-development of explainable AI toolkits and model critique workflows.

7. **Pilot Human-AI teaming systems across multiple scales and domains:** Fund research as well as deployment testbeds at local and global scales in domains where humans and AI collaborate in real time, including disaster response, health, and environmental monitoring, building scalable models for multi-agent systems.

### Training and Capacity Building

This theme focuses on developing the human capital — the skills, knowledge, and organizational structures — needed to create, manage, and participate in convergence projects across all sectors.

1. **Cross-disciplinary training programs:** Develop new academic and practitioner pathways combining AI, participatory science, design, and policy.

2. **Leverage citizen science for computational workforce development:** Integrate computational citizen science into K-12 and continuing education to build STEM, ML/AI and civic literacy skills.

3. **Embed real-time feedback systems via AI across citizen science platforms:** Deploy systems incorporating AI that train and guide users, effectively building user capacity and skills while also improving data quality.

4. **Broaden participation:** Using AI, bring citizen science into familiar digital and physical environments (apps, games, AR) to expand participation.

5. **Incentivize continued dialogue:** To capture the momentum from this visioning workshop and increase capacity, establish mechanisms to sustain knowledge exchange and best practice sharing. This can be accomplished through: (a) annual convergence research summits, (b) dynamic report updates, (c) online community of practice and knowledge exchange, and (d) establishment of an evaluation and metrics working group.

By strategically converging citizen science and computation, America can usher in a new era of scientific innovation, technological leadership, and public engagement. These coordinated efforts promise not only to sustain, but also significantly amplify the United States' global leadership in science and technology, improve public-sector efficiency, and meaningfully engage millions of citizens in the essential scientific questions of our time, while upskilling them on the use of AI. The time to act decisively, through informed policy and targeted investment, is now.

## INTRODUCTION

Scientific progress and technological innovation are the twin engines that powered the American golden age of science in the 20th century. These triumphs, ranging from breakthroughs in space exploration to advances in biotechnology to the personal computer revolution, were not accidental; they emerged from thoughtful investments, strategic public policies, and deliberate support for scientific research and technological innovation. As we reposition ourselves for the coming decades, we must consciously invest in novel strategies and research infrastructures to secure American scientific leadership and national competitiveness.

The time, talent, and expertise of over 300 million Americans represents a largely untapped national resource for competitive advancement. A pivotal pathway to future scientific and technology leadership lies in the convergence of computational and citizen science[2] research. Smartphones, embedded sensors, cloud platforms, and open-source tools developed through computational research have enabled everyday people to collect, contribute, and analyze data alongside professional researchers. At the same time, artificial intelligence (AI), particularly machine learning, deep learning, and large language models (LLMs), has opened powerful new ways of discovering patterns, generating insights, and automating scientific tasks (Fortson et al., 2024). Volunteers can engage with novel systems to advance diverse citizen science goals including data collection, classification, exploration, and interpretation. In several arenas of computational research (e.g., Human Centered Computing, Human-Computer Interaction, and Computer-Aided Cooperative Work), real-world systems provided by citizen science and crowdsourcing enable research in critical areas such as: (1) optimized roles in human-computer teaming specifically when AI assistants are incorporated; (2) development of trustworthy, explainable AI particularly for complex or sensitive problems; and (3) the impact of engagement and motivation of human workers on product outcomes.

Recognized by Congress with the *Crowdsourcing and Citizen Science Act of 2016*[3], open public involvement in science and engineering research through citizen science, crowdsourcing, and other participatory methodologies has clear economic benefits. For example, more than 8,600 citizen scientists volunteering with NOAA's National Marine Sanctuary System contributed time and expertise valued at $1.9 million, equivalent to approximately 40 full-time employees, extending federal capacity while reducing operational costs (Howes, 2017). Similarly, an economic analysis of seven large-scale Zooniverse[4] crowd science projects found that contributions generated between $1.5–$1.55 million in value within just the first 180 days of operation (Sauermann and Franzoni, 2015). In 2023 alone, the Zooniverse platform registered 1.6 million hours of effort contributed by volunteers, which equates to $11.6 million in value at the current national minimum wage. These figures illustrate the substantial fiscal impact of citizen science and underscore its value as a cost-effective strategy for advancing scientific research at national scale. Yet, to keep this vast volunteer corps engaged requires investment in advanced infrastructure, as well as meaningful human and social support structures to facilitate ongoing participation.

---

[2] According to the Crowdsourcing and Citizen Science Act of 2016 (15 U.S. Code § 3724), citizen science is defined as an open collaboration in which individuals or organizations voluntarily participate in the scientific process—such as formulating research questions, collecting or analyzing data, developing technologies, or interpreting results—while crowdsourcing refers to obtaining services, ideas, or content through voluntary contributions from a group, especially via online platforms. In computer science, crowdsourcing can be associated with paying multiple, independent contributors to perform micro-tasks.

[3] Language from HR 6414: Agencies shall: (1) make public and promote such projects to encourage broad participation of consenting participants, and (2) endeavor to make data collected through such projects open and available, in machine readable formats, to the public.

[4] With nearly 3 million volunteers, Zooniverse is the world's largest citizen science platform for data analysis, hosting hundreds of research projects across all disciplines.

Beyond economic value, citizen science contributions are profound and diverse (*see the inset boxes throughout the report for examples of existing highly successful citizen science platforms and projects*), ranging from discoveries of new types of galaxies (Cardamone et al., 2009) to Nobel prize-winning approaches (Koepnick et al., 2019) in protein folding that are accelerating drug discovery and improving public health. Biodiversity monitoring through citizen science has led to the discovery of previously unknown species (Mesaglio et al., 2025). These are more than anecdotal success stories; they demonstrate the enormous untapped potential inherent in public engagement combined with advanced computational capabilities.

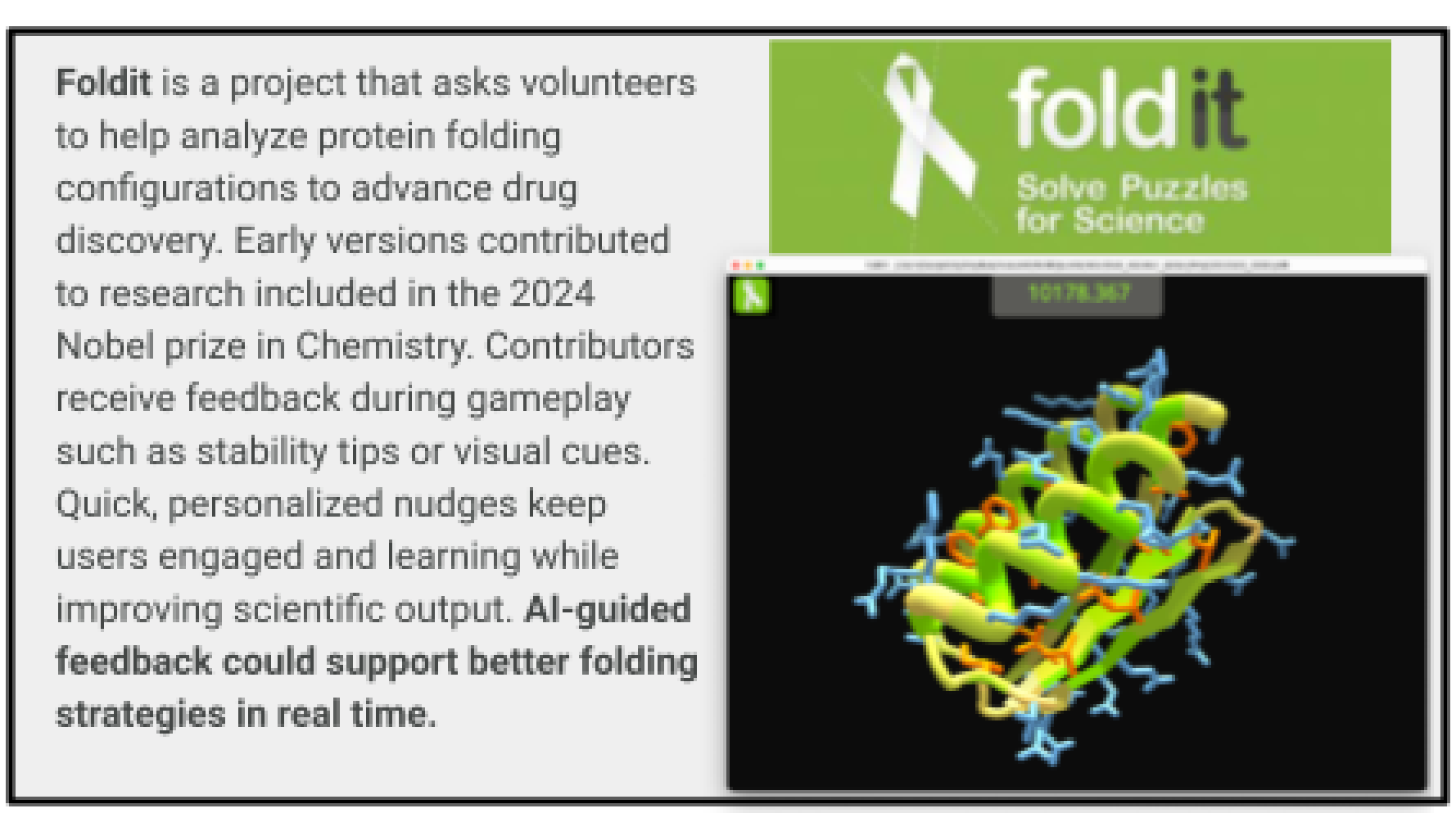

To fully realize this potential, the intentional and strategic convergence of citizen science with the advancing fields of artificial intelligence, machine learning, and computational infrastructure is needed. While each domain has demonstrated independent success, their integration offers exponential benefits. AI can process and interpret vast, complex data sets, including those generated by citizen scientists; in turn, public contributors can enrich machine learning models with local insights, contextual nuance, and edge-case observations that purely automated systems often miss. This symbiosis supports not only more accurate and timely scientific outputs, but also opens up vital computational research such as building trust, access, and transparency into AI systems by embedding them in real-world participatory contexts.

Several state and region-level initiatives already demonstrate how human participation, combined with advanced technology, can deliver measurable public benefits. In Oklahoma, the Mesonet system, a statewide network of automated weather stations, has become a national model for integrating real-time environmental data into forecasting. With AI-enabled anomaly detection and coordination with citizens, it could offer earlier warnings of tornadoes or flash floods, reducing loss of life and property. In Colorado, the DNR's Colorado Corridors Project engages citizen scientists through Zooniverse to map wildlife crossings, working collaboratively with state and federal partners to inform transportation and conservation planning.

Convergence of machine learning for image analysis and human expertise could streamline identification of critical crossing sites and predict emerging animal movement patterns. In Pittsburgh, the Smell Pittsburgh app empowers residents to report odor events, blending community-generated observations with municipal air-quality data to visualize pollution patterns. AI-driven modeling of odor reports, combined with human input, enables predictive alerts about pollution hotspots, directly informing local health warnings.

At the federal level, the National Oceanic and Atmospheric Administration (NOAA) Great Lakes Underwater Videos initiative engages citizen scientists to annotate underwater footage for habitat and invasive species monitoring. By integrating AI-assisted video analysis with human validation, the project enhances both the speed and accuracy of aquatic ecosystem assessments across the region. At the National Aeronautics and Space Administration (NASA), the Eclipsing Binary Patrol project combines machine-learning detection of stellar candidates with volunteer validation to identify thousands of new binary star systems, expanding our understanding of stellar phenomena.

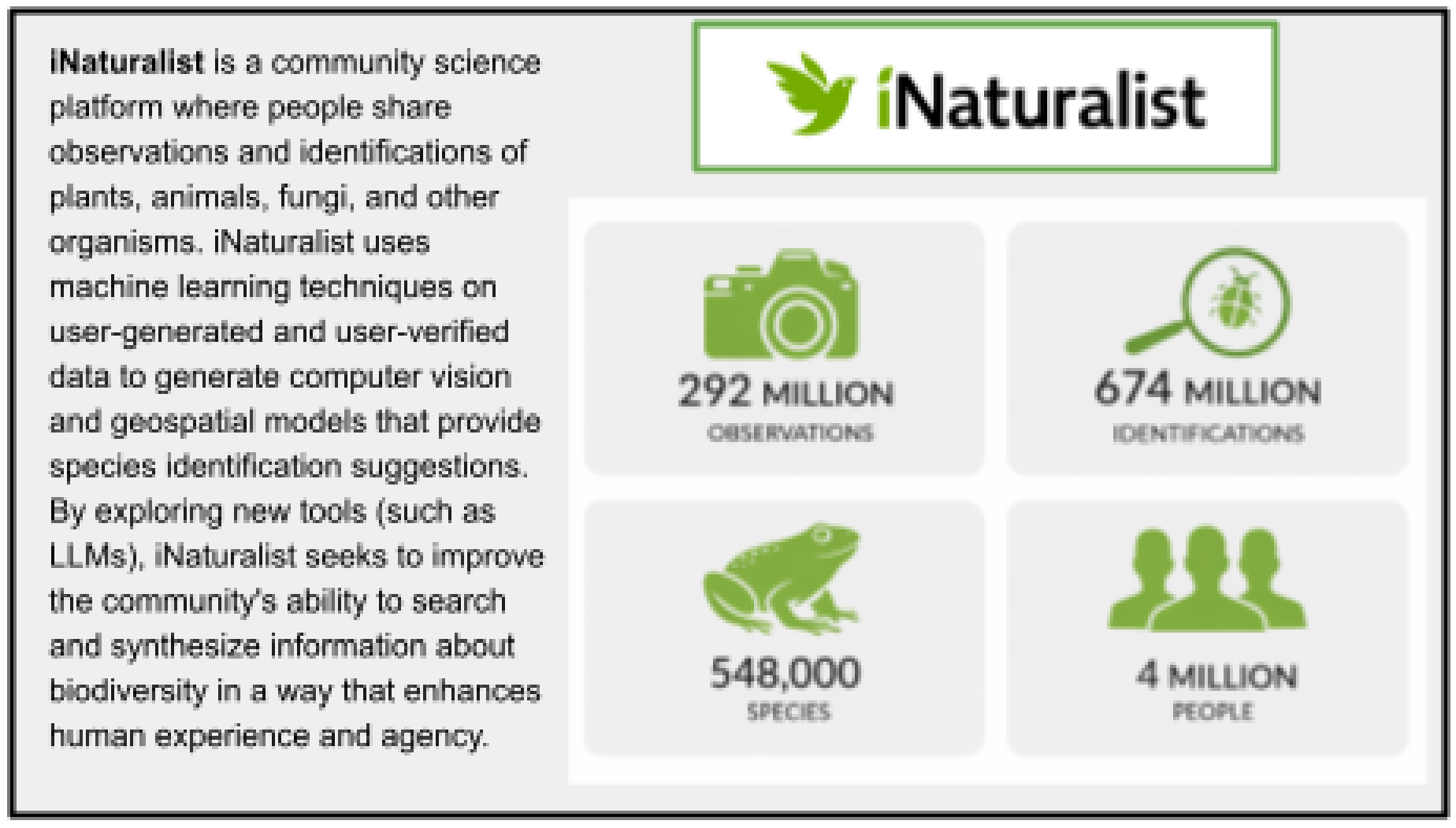

These are just a few projects that demonstrate how hybrid human-AI systems can extend the reach of scientific infrastructure, turning localized observations into actionable insights that improve public safety, environmental stewardship, and scientific discovery. However, they also highlight the need for strategic investment in shared infrastructure, governance frameworks, and research capacity to move from isolated successes to scalable, sustainable systems that operate across domains and jurisdictions.

This report is an outcome of a Computing Community Consortium (CCC) visioning workshop conducted on April 8-9, 2025 in Washington, D.C., and precursor virtual discussions, in which experts across relevant disciplines were brought together to develop a research agenda that brings to fruition the above vision on how humans and machines may team up to solve some of the most pressing scientific problems.

In the following sections, we describe why and where convergence is needed, summarizing the key strategic research directions (called *drivers*) that form actionable pathways for long-term investments in infrastructure, workforce development, and public engagement. We then discuss how these investments in convergence deliver measurable returns for national priorities including national competitiveness, economic value and efficiency, as well as societal benefits such as education and workforce skills development. Substantial detail on each driver is provided in the section on Future Research Directions, leading to the key findings from the workshop. The drivers and findings are then translated into research priorities and recommendations for action. Taken together, the report provides a research roadmap for the convergence of computation and citizen science that culminates in a long-term vision of resilient, adaptive, and equitable systems for human-computer collaboration in tackling 21st-century challenges facing scientific research.

## WHY CONVERGENCE IS NEEDED

The sheer volume of data generated by diverse instruments — from satellites to biological imaging systems — has already created an analysis gap that citizen science is helping to bridge. New technologies are widening this gap still further, producing data at unprecedented scales and levels of complexity, much of it heterogeneous and unstructured. Addressing these challenges requires human cognitive strengths that remain difficult to automate, including contextual interpretation, hierarchical reasoning, problem-solving, and creativity (Rafner et al., 2022). Citizen scientists not only provide these uniquely human capabilities, but also contribute to the collection of new data and the formulation of novel scientific questions, expanding the scope of inquiry in ways computational systems alone cannot.

In Alzheimer's research, the **Stall Catchers** project turns microscopic brain imaging into a global citizen science challenge. Volunteers examine short video clips of blood vessels and label them as either "flowing" or "stalled," generating large-scale annotations that would take researchers years to complete alone. AI models help pre-process and filter the imaging data, while human players contribute fine-grained judgments that remain difficult for machines to achieve reliably. Together, this collaboration accelerates discoveries about blood flow and memory loss, showing how human intuition and AI pattern recognition can complement one another in medical science.

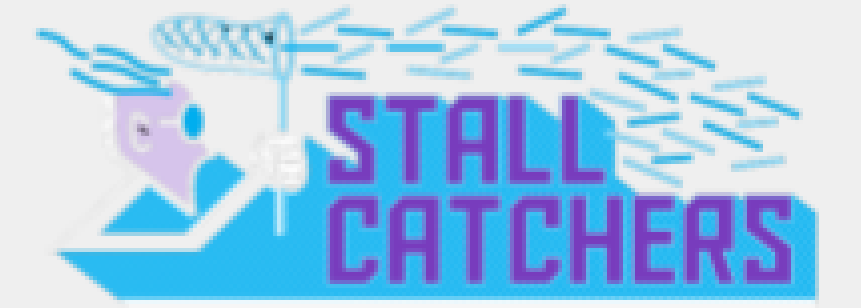

Indeed, the rise of machine intelligence coupled with human intelligence, along with the growth of computational technologies used for data production, modeling, analysis, and deployment, presents a compelling opportunity to scale the use of citizen science. Joined with a receptive scientific community, citizen science at scale can be used to address pressing scientific and societal issues at the state, national, regional, and global levels (Lotfian, Ingensand, & Brovelli, 2021).

Without purposeful convergence, the gap between what our computational systems can produce and what our scientific communities can interpret will continue to widen. Convergence offers a path to close this gap: computation provides speed, scale, and pattern recognition, while citizen scientists contribute contextual insight, local knowledge, and creativity that machines alone cannot replicate. Together, they create a socio-technical system capable of tackling questions that are too large, too complex, or too urgent for either side alone.

## WHERE CONVERGENCE IS NEEDED

Workshop participants defined five *research drivers* that set the strategic priorities for advancing the convergence of citizen science and computational research. Within each driver, we highlight specific **research directions** that translate these priorities into actionable pathways for innovation and investment.

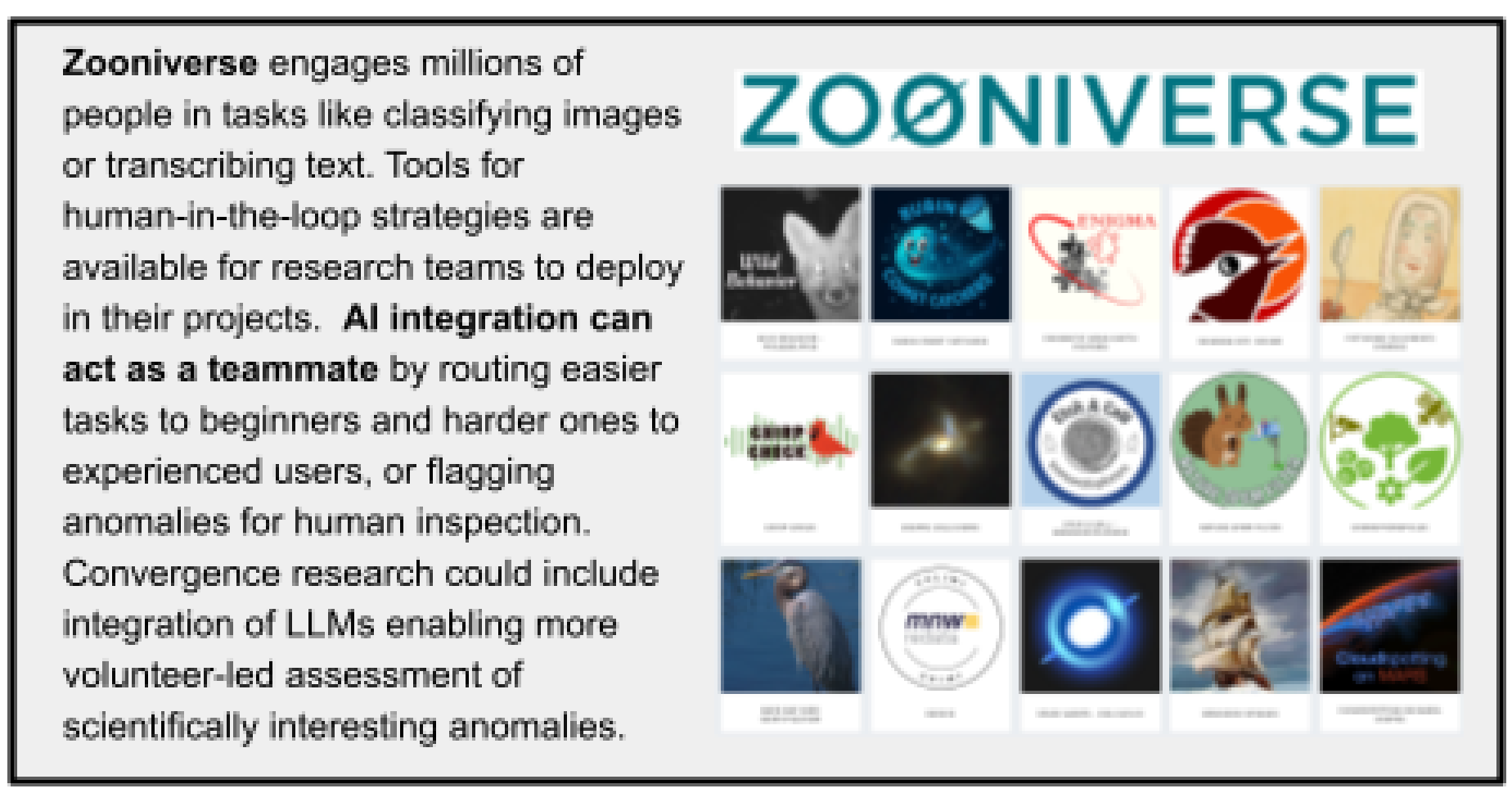

### Driver 1: Human-Computer Teaming: Multi-Agent, Multi-Modal, Multi-Stakeholder Systems

This driver focuses on developing novel ways in which humans and ML/AI can optimally collaborate both in the context of citizen science and in computational research. The aim is to enable multi-agent teams to accelerate scientific discovery while balancing productivity, accuracy, engagement, and education of participants.

1. **Multi-actor:** Optimize how people and machines share tasks so that each contributes its strengths.

2. **Multi-setting:** Bring citizen science into familiar digital and physical environments (apps, games, Augmented Reality (AR)) to broaden participation.

3. **Multi-modal data:**[5] Develop methods for combining and evaluating data from many sources and formats.
4. **Multi-stakeholder:** Align research goals with community values through participatory design.
5. **Evaluation frameworks and metrics:** Create new evaluation metrics that capture the complexity of multi-agent, multi-modal systems.

The convergence of human-computer teaming, advanced ML/AI, and pervasive sensor technology is poised to revolutionize citizen science and augment computational research in areas such as human-centered computing.

## Driver 2: Feedback, Interactivity, and Just-in-Time Delivery

As citizen science expands in scale and impact, especially in the context of our first research driver, there is a growing need for computational systems that support near real-time, personalized, and reciprocal feedback that connects volunteers, scientists, project teams, and societal stakeholders in meaningful ways. To strengthen human-computer teaming, feedback systems must be more dynamic and interactive. Computational systems can play a transformative role by enabling adaptive, explainable, and scalable feedback loops.

1. **Develop Personalized interactions:** Computational systems can provide real-time, personalized guidance to keep volunteers engaged and improve data quality.
2. **Promote Social Engagement:** Systems can support collaboration and peer learning among participants.
3. **Enable Reciprocal Communication:** Systems can enable two-way communication so volunteers feel recognized and informed.
4. **Improve Diagnostic Feedback:** AI can be used to give immediate diagnostic feedback on performance and data reliability.
5. **Expand Societal Communication:** Communicate project outcomes in ways that connect to community priorities and societal impact.
6. **Integrate Iterative Design:** Embed feedback and iteration into projects from the beginning rather than as an afterthought.

[5] Multi-modal refers to combining information from multiple modes of data to obtain a broader, more accurate understanding of an object. For example, training AI on images and their accompanying textual captions provides more accurate models than just using images or textual data alone for training.

Ongoing, meaningful feedback — both personal and social — is key to sustaining participation, improving learning, and ensuring data reliability, and is relevant for both citizen science and computational science purposes.

## Driver 3: Trust, AI, and Citizen Science: Building a Shared Future

To support the Feedback and Human-Computer Teaming drivers, it is critical to establish trustworthy systems. Trust is fundamental for citizen science systems that use ML/AI. To succeed, AI-enabled citizen science must be trustworthy, transparent, and co-governed with communities, ensuring that volunteers feel respected and included. Key directions are:

1. **Advance Governance:** Creating governance models that give communities a meaningful voice in how computational systems and data are used.
2. **Enhance Transparency:** Designing transparent and explainable systems that make AI outputs understandable to non-experts.
3. **Develop Trust Metrics:** Developing metrics that measure not just usability, but also relational aspects of trust, such as respect, perceived accuracy, and shared identity.

Computational systems bring both powerful opportunities and significant challenges to the trust landscape of citizen science. At the same time, citizen science provides opportunities for computational researchers to better understand how trust in AI can be integrated into real-world systems.

## Driver 4: Security, Privacy, and Open Participatory Systems

The first three drivers rely on open participatory systems that should provide both privacy and security. A culture of openness (when appropriate) and respect strengthens credibility, improves societal uptake of results, and empowers more people to contribute to research. At the same time, we must guard against vulnerabilities that are increasing with the use of AI and a system which relies on thousands to millions of citizen science volunteers and their computers to achieve shared goals including decision making and scientific objectives. Priorities include:

1. **Ensure Resilience:** Designing resilient systems that can withstand attacks and maintain integrity.
2. **Defend Against Adversarial Threats:** Defending against malicious attempts to inject false data or manipulate participants.

3. **Strengthen Privacy and Data Governance:** Implementing strong privacy protections and transparent consent processes.

4. **Safeguard Data and Model Sovereignty:** Preserving local control of data and models while still enabling regional, national, and global collaboration.

5. **Balance Openness and Protection:** Balancing openness with safeguards, so systems remain trustworthy but not exploitable.

Convergence of computational and citizen science research is critical to the research in this driver. Again, citizen science projects and platforms provide real-world systems for computational research to study, building open systems that balance accessibility with security and privacy.

## Driver 5: Sustaining Infrastructure across Cyber, Data, Human, and Physical Layers

This driver underpins the previous four as it details the research needed for all the supporting infrastructure. The convergence of technology, particularly personal devices and ML/AI, is accelerating the potential of citizen science. At the same time, the distributed, collaborative, long-term, and contextual nature of citizen science also makes it the most demanding use case for 21st-century scientific infrastructure. Priorities include:

1. **Advance Edge Sensing and Computing:** Developing edge computing and sensing tools that work in low-resource environments.

2. **Develop Lightweight Machine Learning Models:** Creating lightweight AI models that can run on smartphones and low-cost devices.

3. **Build Next-generation Hybrid Connectivity:** Building reliable, hybrid connectivity from local networks to 6G.

4. **Create Decentralized Data Infrastructures:** Designing decentralized, interoperable data systems that make contributions FAIR (Findable, Accessible, Interoperable, Reusable).

5. **Enable Adaptive Participation:** Making platforms adaptive and dynamically scalable to new technologies and user needs.

6. **Ensure Sustainability of Platforms:** Establishing sustainable funding and maintenance models.

7. **Support Development of Practitioners and Participants:** Training practitioners and participants to use advanced tools effectively.

The envisioned scientific infrastructure is one that can support both scientific rigor and public participation at any scale as well as high variability across time, space, and individuals. Both computational and citizen science research needs long-term, flexible infrastructure — both technical and human — that can adapt, scale, and remain sustainable over time.

Figure 1 shows how the drivers enable the convergence of computation and citizen science in service of the cycle of research for which participatory science is an integral part.

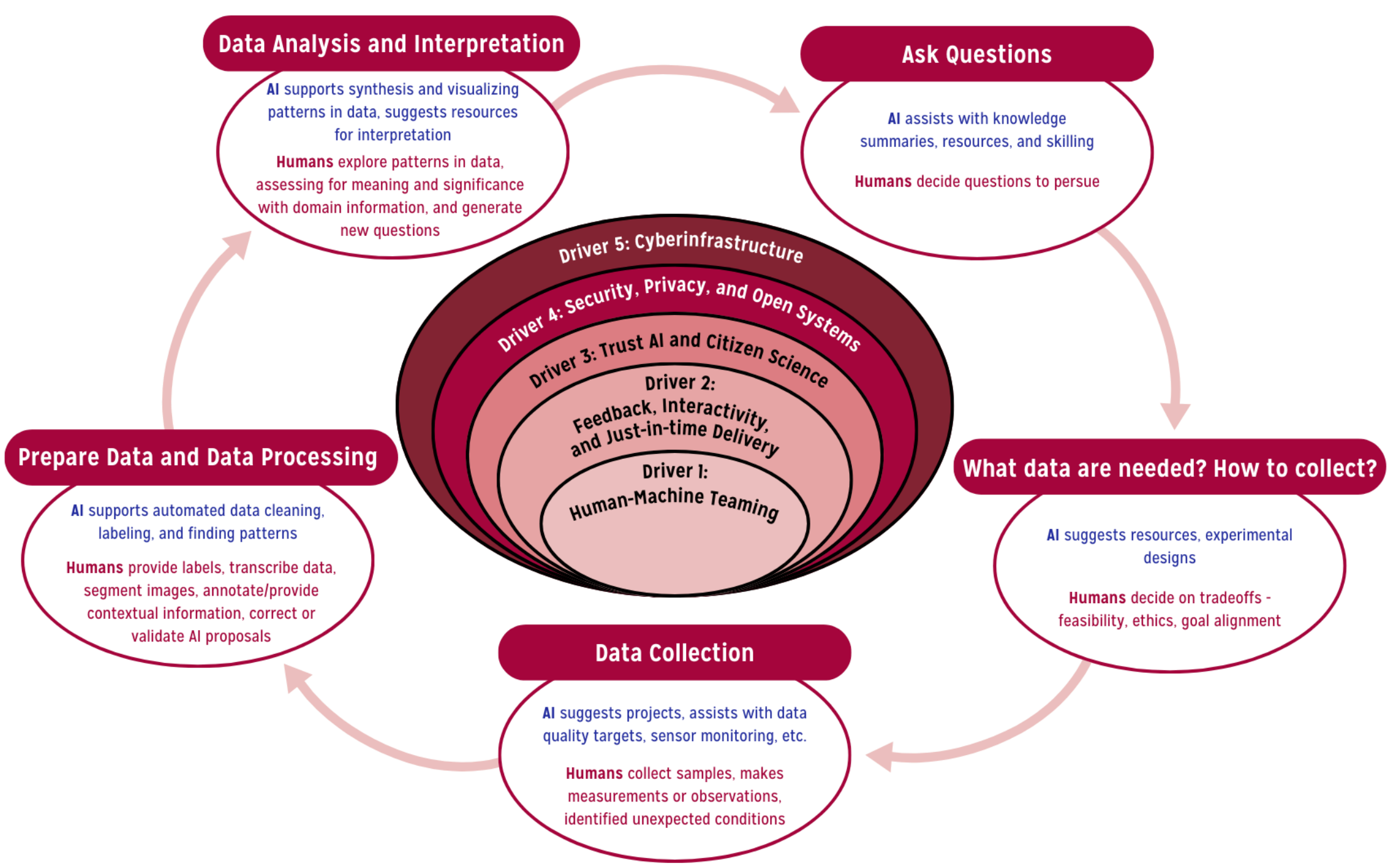

*Figure 1. The convergence of computational and citizen science research, as detailed in the five drivers, will improve the cycle of research. Citizen scientists may contribute to the scientific process at any point in this cycle. Their participation may depend on their motivation, availability, specific skills, background knowledge, and willingness to learn. In addition, the mode of participation can range from dabbling all the way to co-creation of research questions and outcomes. The research cycle shown here highlights several of the ways in which citizen scientists can participate and how computational systems and AI might assist.*

## CONVERGENCE WITH NATIONAL INTEREST

The convergence of citizen science, crowdsourcing, and computational research is not simply a visionary goal — it is a strategic necessity grounded in national interests and aligned with recent federal directives, including M-25-21: *Accelerating Federal Use of AI through Innovation,*

*Governance, and Public Trust,* and M-25-22: *Driving Efficient Acquisition of Artificial Intelligence in Government,* which emphasizes domestic technological leadership, infrastructure modernization, national resilience, and economic competitiveness. These priorities create both a demand and an opportunity for expanded integration of citizen science and AI-enabled research infrastructure into the national innovation ecosystem. The U.S. can lead globally in both technological innovation and participatory governance of science by supporting convergence around the following national priorities:

## Enhancing National Competitiveness

Maintaining scientific leadership is directly linked to national competitiveness. Citizen science and crowdsourcing enhanced by computation (and vice versa) can accelerate scientific discovery, simultaneously supporting rapid innovation across multiple fields. By leveraging the strengths of distributed human intelligence coupled with computational accelerators such as machine learning as well as technology supporting broad deployment and collaboration, the United States can generate faster insights from complex data and stay ahead in critical domains, such as physical sciences, biotechnology, and space surveillance. Convergence also fosters a more agile research ecosystem, allowing institutions to respond more quickly to emerging challenges and opportunities (e.g., monitoring and responding to natural disasters). Moreover, widespread public participation in scientific endeavors cultivates a technically literate population, expanding the pipeline of STEM talent essential for sustaining long-term economic and strategic advantage. (**Relevance**: *M-25-21 § Driving AI Innovation, M-25-22 §1 Ensuring the Government and the Public Benefit from a Competitive American AI Marketplace and §3.c Maximize the Use of American-Made AI* → **Drivers**: *Human-Computer Teaming (1), Sustaining Infrastructure (5)*).

## Economic Value and Government Efficiency

Beyond accelerating innovation, the convergence of citizen science, crowdsourcing, and computational research will yield measurable economic benefits. Citizen science offers a multiplier effect expanding research reach, building public trust, and supporting education, collaboration, and economic opportunity in ways few other strategies can (AAPS, 2025). Integrating distributed human intelligence to improve AI systems through volunteer labeling and data generation can reduce duplication of efforts, streamline workflows, and lower the cost of data collection and analysis across state and federal agencies. This application not only yields direct budgetary savings but also improves the timeliness and accuracy of decision-making in critical areas such as public health, agriculture, and transportation. Additionally, by enabling more intelligent allocation of government resources and increasing the responsiveness of public services, this convergence directly supports the goals of modernizing government operations for greater efficiency and impact. (**Relevance**: *M-25-21 §  Guidance on Federal Use* *of AI, M-25-22 §1 Safeguarding Taxpayer Dollars by Tracking AI Performance and Managing Risk, and § 4.b: Market Research & Planning* → **Drivers**: *Human-Computer Teaming (1), Privacy, Security (4), Sustaining Infrastructure (5)*).

## Workforce Skills Development

As the nation faces growing workforce needs in technology-driven sectors, citizen science and crowdsourced projects, particularly when integrated with AI, are potent avenues for skills development. Citizen scientists, through hands-on involvement in research projects, can acquire essential skills in data literacy, computational thinking, and problem-solving, which are crucial for maintaining America's economic leadership and innovation capability. These experiences can serve as accessible entry points into high-demand STEM careers, particularly in upskilling existing workforces (e.g., industrial maintenance where workers must interpret dashboards and monitor data streams). Embedding real-world computational systems and AI applications into public-facing science initiatives enables scalable, lifelong learning pathways that complement formal education, leading to a more technically capable population and a stronger national workforce pipeline in areas vital to global competitiveness. (**Relevance**: *M-25-21 §Section 2.a.ii.F: Recruit, hire, train, retain, and empower an AI-ready workforce and §3.e: Enabling an AI-Ready Federal Workforce* → **Drivers:** *Feedback, Interactivity (2), Trust (3), Sustaining Infrastructure (5)*).

## Computation and Infrastructure Development

Realizing the full potential of citizen science, crowdsourcing, and computational convergence necessitates significant research infrastructure investments beyond traditional computing hardware and software. An integrated, adaptive computational infrastructure is needed to manage and analyze large-scale, heterogeneous data sets and explore novel scientific questions. Furthermore, this infrastructure needs to comply with the Open, Public, Electronic and Necessary (OPEN) Government Data Act. Specifically, this infrastructure encompasses networking technologies, sensors, edge computing, AI models, cybersecurity measures, and accessible software solutions. It must be robust enough to address diverse local, state, national, and even global scientific challenges, including disaster response, health research, urban planning, and resource management. (**Relevance**: *M-25-21 §2.a.ii.A: Develop AI-enabling infrastructure across the AI lifecycle, §2.a.ii.E: Develop operations and infrastructure to manage risks, §Section 3.b: Update Agency Policies (on IT infrastructure)* → **Drivers***: Human-Computer Teaming (1), Privacy, Security (4), Sustaining Infrastructure (5)*).

## Harnessing Public Insight and Data

The inherent value of citizen science and crowdsourcing is amplified through the effective integration of AI systems that can synthesize and analyze large, heterogeneous data sets more efficiently than human-only teams. Engaging the broader public in scientific inquiry allows research communities to capitalize on the full range of perspectives and localized knowledge that traditional scientific methods often miss. Critically, citizen science and crowdsourcing promote the sharing, reuse, and integration of public and open datasets, consistent with federal directives emphasizing data sharing and reuse. Additionally, emphasis on documentation and transparency ensures that citizen science and crowdsourcing projects produce robust scientific outcomes and uphold rigorous standards of transparency, allowing project participants clear insight into how their contributions inform results. Incorporating AI in all of these aspects will help lower the barrier to participation for a wider range of the public. (**Relevance**: *M-25-21 2.b: Sharing of Agency Data and AI Assets, 2.d: Effective Federal Procurement of AI – Maximizing the Value of Data for AI, §4.b.ii: Complete AI Impact Assessment (documentation and transparency) and M-25-22 §3.g: Contribute to a Shared Repository of Best Practices, §Section 4.c.i: AI Use Transparency Requirements, and §4.d: Selection and Award – Testing and Evaluation (building trust through data transparency →* **Drivers:** *Human-Computer Teaming (1), Feedback (2), Trust (3), Privacy, Security (4)*).

## Empowering Agency and All Viewpoints

Convergence initiatives also offer social benefits, particularly through increased public agency. When individuals see their data, insights, and research questions meaningfully reflected in scientific outcomes, it validates their role as active contributors rather than passive recipients of scientific expertise. AI tools can further personalize participation, allowing individuals to contribute in ways that align with their skills, interests, and experiences. This empowerment fosters a sense of ownership and civic identity, reinforcing that science is a collective endeavor shaped not just by experts but everyone. Additionally, citizen science opens avenues for meaningful public engagement in scientific discourse, providing opportunities for education across all age groups, significantly enhancing STEM learning, especially within K-12 environments. This alignment with public participation and education resonates clearly with the transparency and accountability directives, which emphasize fostering public trust in governmental AI usage. (**Relevance**: *M-25-21 § Fostering Public Trust in Federal Use of AI, §4.b.ii.C: AI Impact Assessments – Public impact, civil rights, and transparency and M-25-22 §1: Promoting Effective AI Acquisition with Cross-Functional Engagement, §3.d: Protect Privacy, § 4.c.i: AI Use Transparency Requirements →* **Drivers**: *Feedback (2), Trust (3), Privacy, Security (4)*).

### Natural Disasters and Response

Citizen scientists with mobile devices, sensors, and local knowledge can collect and validate critical real-time data during flood events, wildfires, earthquakes, and hurricanes—often faster and with more granularity than official channels. AI-enhanced citizen science and crowdsourced systems can then rapidly analyze this influx of data to provide early warnings, optimize emergency response logistics, and support post-disaster assessments. Such developments can enhance public safety, reduce response times, and mitigate disasters' long-term economic and social costs (Simmons et al., 2022). Ensuring transparency in data sources and decision models improves outcomes and builds public trust, which is critical when lives and livelihoods are at stake. Investing in the collaboration of citizen science, computational systems, and AI for disaster readiness and recovery is a strategic imperative for national resilience. (**Relevance**: *M-25-21 §2.a.i: Agency AI Strategies – Current and Planned AI Use Cases, §4.b.iii: Ongoing Monitoring for Performance and Potential Adverse Impacts and M-25-22 §4.a.ii: Determining the Use of High-Impact AI, §4.d.i: Testing and Evaluation (including real-world use conditions* → **Drivers:** *Human-Computer Teaming (1), Trust (3), Sustaining Infrastructure (5)*).

## FUTURE RESEARCH DIRECTIONS

The visioning exercise proceeded in two rounds, first online with a broad audience and then in person by invitation, as presented in the Appendix. What follows is a summary of the findings across all of the workshops, the opportunities and challenges specific to each of the drivers, and recommended actions to be taken if the United States is to remain an innovator in scientific research and development.

### Driver 1: Human-Machine Teaming: Multi-Agent, Multi-Modal, Multi-Stakeholder Systems

New challenges in knowledge production are driven by the increasing amounts of complex data captured across a wide range of spatial and temporal scales, as well as across a variety of sources and formats. To take advantage of the depth and richness in these data sets, there is a need to combine information across different domains, scales, and data modalities. In tackling these challenges, ever more sophisticated digital approaches, including ML/AI, are being applied within scientific domains. Yet even the most advanced computational approaches suffer from issues such as data drift, model generalizability, transparency, and contextual misinterpretation of AI model outputs. For example, models trained to detect specific tumor types from one scanner can yield erroneous outputs when applied to images taken from a slightly different scanner.

As such, there is still a demonstrated need for human-in-the-loop[6] strategies afforded by judicious teaming of humans and computers. While capable of learning patterns beyond human capacity, AI systems can improve by learning from human analysis, referring uncertain cases to humans for help when needed. Human-machine teams can leverage human knowledge about the world to make sense of anomalies in the data that machines may miss. These abilities are particularly important in identifying potential biases inherent in the AI system that can further exacerbate inaccurate solutions and misrepresentations of society. Human perception and intuition can quickly find solutions to some optimization problems that elude machines, which then serve as learning examples for improving the AI. However, human attention is a scarce resource that needs to be carefully deployed, and the growth in data may soon outpace the growth in engagement. Improvements in AI can help mitigate these issues and others by learning to anticipate human needs and requests thus making the most of human attention.

Citizen science offers a real-world setting to test unconstrained and widely relevant human-computer teaming systems. Headway has been made by many projects and platforms such as iNaturalist and Zooniverse (see Fortson et al., 2024). In addition, the human–machine collaboration model builds on decades of experience with volunteer computing systems, which pioneered large-scale human participation in distributed computation and addressed challenges of heterogeneity, trust, validation, and sustained engagement that remain central to today's AI-enabled participatory systems (Beberg et al., 2009 & Taufer et al., 2006). Moving forward, the challenges facing human-computer teaming within scientific knowledge production are strikingly similar to those faced by the U.S. military in implementing multi-domain operation (MDO) strategies detailed in the recent National Academy of Science and Engineering Report "Human-AI Teaming: State-of-the-Art and Research Needs (2022)" [hereafter NAS 2022].

In addressing these challenges, human-computer teaming research will open a vast range of citizen science opportunities that can take advantage of the explosive growth in AI, sensor development, and computation (e.g., cloud services, edge computing). These developments will usher in a new paradigm for citizen science facilitated by a system that optimizes contributions from both human and machine agents. These multi-agent teams can interact across the full range of data modalities in flexible, open-ended, and user-driven ways. Supported by AI agents, citizen science tasks can be integrated into online and physical spaces people already frequent. This integration would harness project participants' implicit, experience-based expertise and align contributions with the specific data modalities being used. Building on the experience from numerous successful citizen science projects coupled with new research, we can push the frontier to allow new forms of human-AI interaction that maximize synergy. To fully enable research that is both citizen-driven and citizen-monitored, this vision will require mechanisms to align the goals of institutions, AI agents, and community members, and developing models with these aligned values. Five research directions have been identified:

**1. Multi-actor: Optimizing the interaction between human and artificial agents in citizen science projects, with a focus on efficient resource allocation and knowledge integration.** Research is needed on methods to facilitate a synergistic relationship between human and AI contributors within citizen science projects and cost-benefit analyses on optimal relationship outcomes. Specifically, as noted in NAS 2022 (pg 17), "descriptive models of human-AI team performance need to be extended into computational models that can predict the relative value of teaming compositions, processes, knowledge structures, interface mechanisms" etc. In citizen science, these models can provide a guide for strategically combining the unique strengths of both humans and AI to enhance the efficiency of data collection and analysis by prioritization and task routing among different agents/entities/modalities. For example, an AI-agent can assist with pre-filtering or pre-labeling data to help volunteers handle complex tasks, potentially routing specific data to individuals with demonstrated skill or capacity. However, such approaches should consider what losses are evident (e.g., efficiency, retention of participants, removal of novel perspectives, etc.) when exploiting prioritized expertise versus exploring a broader set of human or AI contributors.

[6] Human-in-the-loop refers to the need for one or more humans to validate and/or correct the output of an AI algorithm.

Research is also needed on ways to achieve synergies from complementary capabilities. For instance, how might we leverage AI explainability to guide humans to build on AI outputs or conversely, might we design better explainability by feeding human sensemaking to AI models? How might we investigate the best combination of human and machine intelligence to identify scientifically meaningful anomalies in the vast, complex data sets that define 21st-century science? We also need to understand the cognitive difficulty placed on humans in these complex "multi-everything" scenarios and determine to what extent AI agents exacerbate human frustrations or can mitigate lack of expertise. In addition, research is needed on how participants themselves can use or be supported by AI to solve scientific problems using citizen-science generated data and information.

**2. Multi-modal data: Improved methods of data integration and valuation in citizen science, particularly in multi-modal contexts.**

Most existing data sets are limited in modalities (for example, only one or two of the following: text, images, wavelength-specific sensor outputs, audio, etc.). While tools for interaction with these data may be ubiquitous, they are often developed in silos and not accessible or interoperable across user communities with expertise in different modalities. And citizen scientists themselves can contribute to these data sets, collecting or analyzing data across different modes, typically without tools that enable sense-making across modalities.

Powerful AI tools capable of handling multi-modal inputs and outputs are becoming a reality, exemplified by recent releases like Google's Geospatial Reasoning. Further computational research in data integration and fusion across modalities promises exciting breakthroughs in research such as biodiversity and natural resource management, assessing geo-hazard predictions, or precision medicine. Incorporating human-in-the-loop strategies facilitated by citizen science offers efficient and scalable solutions to address the inherent challenges in analyzing these larger, more complex, multi-modal data sets including determination of cross-modal verification strategies which often require human sense-making. This would enable evaluation of the relative strengths and weaknesses of different modalities of data, which could then inform future data gathering and storage, as well as modeling priorities.

**Future Scenario: AI-Assisted Tornado Response.** In Missouri, a tornado strikes and volunteers mobilize alongside autonomous drones streaming imagery and sensor data. An AI coordination system fuses human reports with drone feeds, flagging survivors and routing responders in real time. Cloud-edge computing and multi-agent planning enable rapid, life-saving decisions. What once took days—damage assessment and rescue—happens in hours through human-computer collaboration.

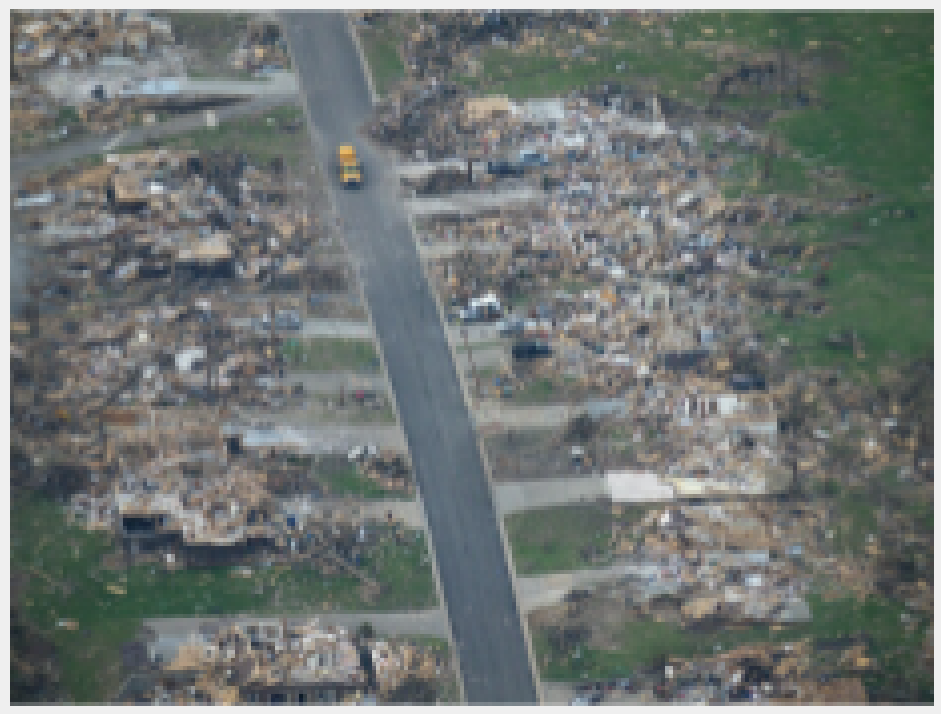

Credit: Bob Webster (CC license)

Research is also needed on methods for assessing the value of data within citizen science projects. Data have a human, environmental, and infrastructural cost that must be balanced with the value of the information they contain for effective data management and decision-making. Facets of the data valuation process need to be captured and maintained longitudinally to support long-term decision-making, metricizing, and benchmarking. However, data value is dynamic and context-sensitive. For instance, we need a better understanding of how collecting new data points adds value to a corpus, a problem that increases in complexity with additional data modalities.

Finally, research is also needed into how to improve data quality submitted by citizen scientists, in particular as new collection and analysis modalities become increasingly accessible. The better the quality and variety of data submitted by humans, the better the training data are for AI. How might we maximize the rigor and reproducibility of integrated human-machine citizen-science projects? How will we address challenges in crowdsourced data provenance, attribution, bias, integrity, and verification?  For example, citizen science combined with augmented reality can improve data quality and quantity by helping to accurately target optimal data collection locations, including natural environment landmarks or locating animals from drone imagery.

**3. Multi-setting: Strategies for seamlessly embedding citizen science activities into everyday digital and physical environments.**

Existing multi-modal systems are beginning to blend physical and digital realms (such as augmented or virtual reality representations) and enable real-time interactions between different human and AI agents. Research is needed on ways that AI can integrate citizen science into the daily lives of potential participants by using their existing engagement in different physical and digital spaces. This approach recognizes the potential of social media platforms, gaming environments, and neighborhood or community hubs to host citizen science activities, thereby tapping into a broader audience and a wider range of expertise and lived experience. Such integration can make participation a natural extension of people's routines, enhancing both the scale and the relevance of citizen science contributions. However, with scale comes noise and potential bias, further research is required into methodologies to glean useful information at larger multi-setting scales and across broader communities, while recognizing that what seems like noise may sometimes be a useful data signal.

Research is also needed on ways to benefit from the physical settings of human participants as well as corresponding digital twins[7]. Approaches are needed for identifying the full range and interdependence of expertise within communities, such as local knowledge of places to support routing of questions to participants who would find them relevant or easy to answer. Furthermore, e.g., using AI on mobile devices to ask participants what other data might be collected in a given location could reduce costs by combining priorities. Research is also needed in situating citizen science in augmented or virtual reality technologies, to both enable extensions of people's local physical knowledge into virtual captures of other, similar environments, as well as exploration of environments not naturally scaled to human interactions (e.g., molecular structures or galaxies).

**4. Multi-stakeholder: Aligning scientific objectives with community needs and values in citizen science projects.**

Citizen science projects must be aligned with the needs and values of the communities they involve. Research is needed to identify effective participatory approaches that foster collaboration between scientists and community members in the design and implementation of projects. How can we use emergent technologies such as generative AI to facilitate the co-creation of goals that center both scientific and community needs? How do we enable transparency around current and future (even hypothetical) use of data across heterogeneous communities, providing mechanisms for opt-in/out for different uses of user-contributed data?

---

[7] A digital twin is a digital replica or simulation of a physical object or system that enables virtual interactions including exploration, manipulation, and augmentation for training purposes or to determine viable ways in which to physically improve the system. An example is a digital twin of an airplane where electrical engineers can make adjustments to the design of certain cockpit functionalities and digitally explore the consequences.

What do consent processes look like for more passive forms of data collection, where people may or may not be aware of how their data are being used? Development of approaches that emphasize transparency and ethical considerations can promote trust and mutual benefit in citizen science partnerships.

**5. Creating innovative evaluation frameworks and metrics for citizen science projects, particularly those involving multi-agent systems and diverse data modalities.**

New approaches are needed to assess the impact and effectiveness of citizen science projects that involve complex interactions between humans and AI across a wide range of data types. Focusing on a single outcome (e.g., quantity of data collected or analyzed) will yield suboptimal designs. Metrics are needed to assess the added value and cost of presenting multiple data modalities in a project as well as metrics to compare purely human, purely machine-based, or ensembles of multi-agent workflows. Evaluation approaches are needed for complex projects that help identify errors or improvements related to specific system components. New approaches are also needed to assess alignment across different stakeholders.

## Driver 2: Feedback, Interactivity, and Just-in-Time Delivery

A second research driver is enhancing support for feedback processes as a vital element of citizen science projects. As citizen science expands in scale and impact, especially in the context of our first research driver, there is a growing need for computational systems that support near real-time, personalized, and reciprocal feedback that connects volunteers, scientists, project teams, and societal stakeholders in meaningful ways. Current systems often overlook feedback or treat it as an afterthought, which forgoes opportunities for improved participation, compromises data quality, and limits scientific progress. Computational systems can play a transformative role in addressing these limitations by enabling adaptive, explainable, and scalable feedback loops. These systems can improve user learning and motivation, support iterative system design, and foster trust across stakeholders, all while accelerating scientific outcomes. We call for research to develop infrastructures where feedback mechanisms are embedded from the start and evolve alongside the system and its users. Six areas for research were identified:

**1. Personalized Interactions: Delivering adaptive, real-time feedback that sustains participant engagement and strengthens scientific reliability.**

Research is needed on how to tailor feedback from platforms to individual volunteers — and the human/computation interface more broadly — to enhance learning, insight, motivation, trust, and sustained participation. Research needs to consider how these goals interact, e.g., applying computational tools to enhance productivity and retention may reduce opportunities for learning (Pankiv & Kloetzer, 2024); volunteers may be demotivated or turned off by technologies that automate interactions or that have significant environmental impacts.

Questions include, for example, how might we vary feedback by modality, tone, and frequency; how might we diagnose a volunteer's learning progress; how to ensure that engagement with the project protects volunteer well-being; and how might we personalize task assignments and support to match a participant's evolving skills and goals, for instance, by using system capabilities to guide volunteers towards more advanced research interactions. In all of these interactions, AI tools such as LLMs in particular could facilitate adaptation to individual volunteers, such as those with different backgrounds, learning styles or comfort levels and enabling multilingual access. As such, research is needed into the novel AI/ML models needed to facilitate dynamic and interactive feedback systems

**2. Social Engagement: Designing collaborative dynamics that foster peer learning, prosocial behavior, and resilient volunteer communities.**

Studies should explore how computational systems can facilitate constructive peer engagement through collaborative, social, or even competitive dynamics. Designing systems that encourage collaboration, knowledge sharing, reuse of others' contributions (Malone et al., 2017), peer learning, prosocial behavior, and community building, while minimizing or preventing exclusionary or harmful conduct, is a core challenge.

**Future Scenario: Real-Time Learning in Local Labs.** In Michigan, students use low-cost sensors to map air quality around their school. An AI guide provides instant feedback, correcting errors and sparking new questions. Results are plotted on shared maps, linking classrooms statewide and connecting students with researchers. This real-time learning boosts engagement, builds STEM skills, and turns local data into community action.

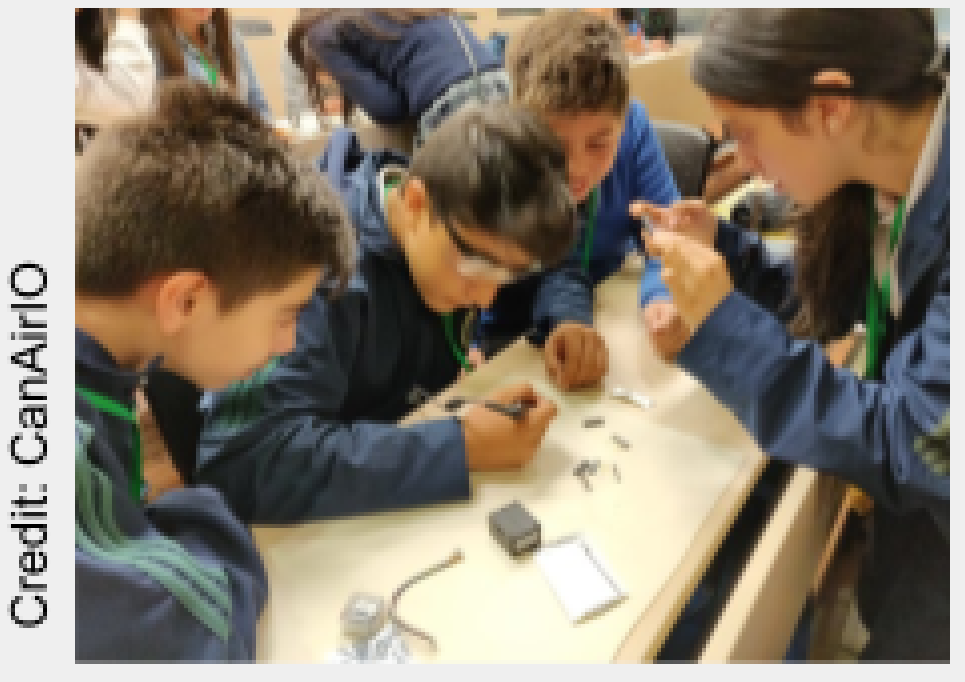

Credit: CanAirIO

**3. Reciprocal Communication: Building transparent two-way exchanges that align participant recognition with scientific accountability.**

We need to investigate how digital platforms can enable reciprocal, transparent, effective, and trust-building communication among volunteers and professional researchers. Systems could help summarize progress, clarify goals, and support two-way exchanges that make volunteers feel recognized and rewarded for their contributions and scientists more informed, while not diminishing the personal connection that many participants value. At the same time, adding AI to workflows may require new training models for volunteers so they understand the complementary roles of volunteers and AI, and can continue to contribute effectively.

**4. Diagnostic Feedback: Monitoring and reporting performance to enhance data quality and participant confidence.**

Systems also must deliver feedback to scientists and developers about system performance, data reliability, and user engagement. This includes research on how to detect design flaws, monitor data reliability, automate reporting, validate the accuracy of automated responses or guidance given to volunteers, and support decision-making by the research team.

**5. Societal Communication: Linking individual contributions to broader social and scientific impacts to expand relevance and trust.**

Future systems should support communication with a broad range of community stakeholders, including educators, policymakers, and different sectors of the public. This vision is grounded in responsible and accountable research practices, ensuring that all voices are heard in shaping scientific agendas and outcomes. The vision includes creating mechanisms for soliciting public input that are accessible to and understandable by all interested parties, developing tools to visualize project impacts, not only in scientific terms but also in relation to societal interests, and building strategies for connecting scientific and broader societal goals in a way that builds transparency and trust. By embedding principles of responsible research and collaborative governance, science teams can ensure that technological and scientific advancements reflect shared priorities and deliver benefits across communities.

**6. IIterative Design: Embedding inclusive, agile co-creation methods to sustain participation, and adapt platforms over time.**

A final research priority is the role of design itself, particularly when AI is incorporated into workflows. Designers should be integrated from the start of project development to help embed feedback loops that support iteration, onboarding, and adaptation. Research (and sharing of best practices) is needed on appropriate design methodologies for citizen science projects and platforms that support inclusive, agile, and co-creative socio-technical system development.

## Driver 3: Trust, AI, and Citizen Science: Building a Shared Future

While computational systems can expand analytical capacity and support new forms of discovery, they simultaneously introduce complex trust dynamics that must be carefully navigated. The challenge is multifaceted: people must trust not only the information that systems provide and their credibility as an expert or synthesizer, but also the broader infrastructure's ability to recognize contributions, reward insights, and protect identities.

The stakes are particularly high given computation's potential to either democratize or further stratify scientific participation. Opaque algorithms, inaccessible interfaces, and lack of transparency risk eroding trust, especially among communities historically excluded from scientific processes. When trust fails in one dimension, whether in the system itself or the surrounding infrastructure, it can undermine confidence across the entire ecosystem, potentially limiting participation to those already privileged within traditional scientific networks.

Workshop participants identified an urgent need for research and infrastructure development that treats trust as foundational throughout the system and project lifecycle. This approach requires developing explainable computational tools that clearly communicate aspects such as computational uncertainty. Only by addressing trust holistically can we unlock the potential to enhance both the integrity and inclusiveness of science. We expand on these ideas and identified the following key areas of research:

**1. Governance: Embedding participatory authority over computational tools through transparent stewardship, and equitable decision-making.**

A first research challenge is understanding how to build trust through participatory governance. New technology should not be assumed to be beneficial or welcomed by all stakeholders. Research should first explore how communities can collectively determine when computation adds value versus when it may undermine their existing practices. Central to building trust is developing processes that demonstrably respond to community priorities about technology use, not just technical possibilities. At the same time, we must be mindful of the social complexities inherent in participatory governance especially in the context of ubiquitous availability of commercial LLMs and data use/reuse. Research on effective governance structures with options for participants is needed.

**Future Scenario: Crowd-Powered Meteoroid Surveillance.** Across Texas and beyond, amateur astronomers track Near Earth Objects and feed data to an AI platform that calculates precise orbits from the distributed data. Within minutes, Earth intercept times and locations are determined along with potential impact parameters. Details are immediately shared with NASA's Meteoroid Environment Office, where alerts are issued based on impactor size. Citizen observations, combined with AI validation, form a powerful distributed sensor network. This collaboration builds trust, strengthening Planetary Defense capabilities at the same time as contributing to scientific understanding of our solar system.

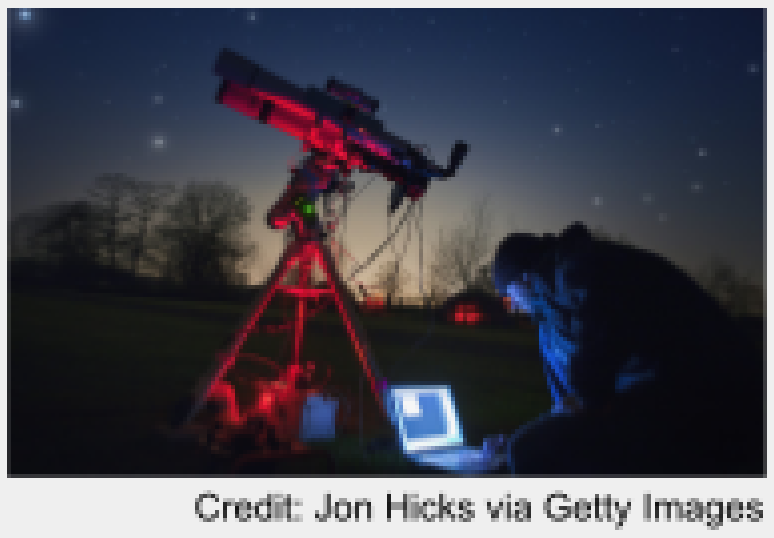

Credit: Jon Hicks via Getty Images

When technology integration is deemed appropriate, trust emerges through transparent controls that allow citizens choices on how their data is licensed, shared, and used over time, with clear mechanisms to update consent, and receive alerts about new uses. This trust-centered approach extends beyond data stewardship to examine how citizens can meaningfully influence AI model goals and algorithmic behavior in ways that feel authentic rather than performative. Accordingly, research is needed on trust-building mechanisms within existing governance structures while also exploring how to design new participatory frameworks for technology decision-making. This direction includes developing trustworthy processes for citizen input in publicly funded initiatives, and creating accountability measures for private entities using citizen-generated data. Key trust considerations involve ensuring citizen voices carry actual weight in both adoption decisions and commercial applications, establishing clear transparency standards that citizens can verify, and designing governance systems where proprietary interests don't override participant agency.

**2. Transparency: Making computational processes understandable, relatable, and accountable to support informed participation.**

Research is needed on how to create citizen science platforms and computational tools that not only convey decisions in understandable ways but also actively prevent blind acceptance of outputs. To achieve this goal requires developing systems that use plain language, relatable visualizations, and uncertainty-aware outputs while incorporating built-in safeguards that encourage critical evaluation. Key questions include how to design interpretable models that adapt to volunteer's data literacy levels, how to implement guardrails that keep recommendations within appropriate bounds, and how to create interfaces that prompt users to question and verify suggestions rather than accept them without scrutiny.

Operational trust also depends on robust provenance mechanisms that capture how data, models, human contributions, and computational workflows evolve over time. Provenance enables auditability, reproducibility, attribution, and accountability, critical requirements for trustworthy participatory AI systems. Research is needed on how to develop scalable, provenance frameworks for participatory AI systems. This includes research on lightweight, privacy-aware provenance capture that spans human actions, AI model evolution, and distributed execution, enabling reproducibility, auditing, and long-term reuse at the national scale.

**3. Trust Metrics: Developing diagnostic frameworks to assess relational trust, power dynamics, and perceived expertise across networks.**

New research is needed to create comprehensive metrics that assess the health and dynamics of collaborative relationships in citizen science, which is key to support the design of trustworthy AI interfaces and communication methods. This research direction focuses on measuring and studying the underlying social relationships and collaborative dynamics that emerge over time, as well as how trust affects them.

The central goal of this research direction is to develop and validate new frameworks for measuring mutual trust and its direct impact on efficacy and creativity across the entire network of volunteers, scientists, platforms, and computational tools. This work moves beyond subjective assessments to ask a more concrete question: How can we reliably quantify the trust within these complex collaborations? To answer this, we need to develop frameworks that measure trust by evaluating its foundational elements, including perceived accuracy, demonstrated performance, personal familiarity, or shared identity. Measuring these specific elements is critical because it provides a diagnostic tool. It allows us to assess not only *if* trust exists, but on what basis it is built, thereby making it possible to diagnose issues and intentionally cultivate the type of trust required for both efficient collaboration and creative scientific breakthroughs.

The metrics for measuring trust must capture relational outcomes that extend far beyond interface usability or information comprehension. For example, do participants feel their expertise is genuinely valued? Do scientists demonstrate trust in volunteer contributions through their actions, not just their communication? How do power dynamics and bias shape collaborative effectiveness? While systems themselves cannot experience trust, they can be designed to embody trust through their operations by incorporating participant feedback, maintaining transparency, and demonstrating respect for contributor expertise. Novel frameworks should track how these relationship dynamics evolve over time and guide interventions that strengthen the social fabric of citizen science communities, creating resilient and equitable collaborations that can adapt and innovate together.

## Driver 4: Security, Privacy, and Open Participatory Systems

As government agencies and other organizations increasingly rely on the results of scientific research and monitoring that incorporates citizen science and crowdsourced approaches, we will need to identify, understand, track, and mitigate for the possible cybersecurity, cognitive security, and AI adversarial security threats in these systems. Research is needed to characterize and develop strategies to combat computer-mediated manipulation of citizen-science-volunteer behavior, as well as the potential spread and impact of false data through citizen science platforms, applications, and observations (video, text, speech). Five key areas for research are identified:

**1. Cybersecurity and Resilience: Protecting large-scale collaborative platforms from threats while preserving openness and accessibility.**

In open citizen science platforms, robust cybersecurity and system resilience form the backbone of trustworthy operations. These projects often involve a heterogeneous array of user devices (e.g., mobile data collection apps, edge devices, low-cost sensors, drones, etc.), and can scale to tens of thousands or even millions of participants, creating an extensive attack surface for cyber threats. This raises fundamental research questions, including: what are the specific vulnerabilities and their corresponding attack paths across the citizen science or crowdsourcing pipeline (from user devices to cloud infrastructure), and how might each component of the system, including hardware and software, be fortified against intrusions or failures without deterring public participation or slowing down science and innovation? Ensuring resilience means the system must maintain integrity and availability even under attack or failure through measures such as real-time anomaly detection, data validation, and fault-tolerant design. This theme is strategically important because a single breach or data integrity failure can erode public trust, compromise sensitive data, and derail scientific outcomes. Progress in cybersecurity and resilience will safeguard the reliability of citizen science and crowdsourcing methodologies, enabling wider adoption and the safe realization of its scientific and societal benefits at scale.

**2. Adversarial Threat Mitigation: Developing automated defenses against manipulation, disinformation, and data poisoning.**

Citizen science systems face active adversarial threats ranging from subtle data poisoning to coordinated disinformation attacks. Malicious actors may operate individually or as a team to contribute falsified or manipulated data and even attempt to sway volunteers' behavior with false information and misdirection. These adversarial interventions can distort scientific analyses and demoralize volunteers if not rapidly detected and mitigated. Key research questions include: (1) how might we reliably identify manipulated or fake data amidst large, noisy, and multimodal datasets; (2) how might we distinguish adversarial attacks from natural errors that might threaten the integrity of AI models trained on crowdsourced data; and (3) how might we develop AI training pipelines that are robust against such threats? Researchers are also investigating how might we detect human-curated versus AI-generated manipulation, and how might we profile, preempt, and counter attackers' tactics. Strategically, advancing adversarial threat detection is essential to preserve the integrity of citizen science and crowdsourced data and downstream models.

We also need to characterize what types of adversaries — from individual malicious insiders and hobbyist "trolls" to organized cybercriminal or state-sponsored actors — are likely to target citizen science and crowdsourcing systems, what are their attack vectors and motivations, and how can mapping these threat actors inform the design of security measures and policies to protect such open collaboration platforms. Without robust defenses, projects could become targets of adversarial campaigns or data poisoning that undermine public trust and scientific credibility. Conversely, effective mitigation of adversarial threats will maintain data quality and confidence, allowing citizen science to inform decision-making reliably.

**3. Privacy and Data Governance: Safeguarding sensitive participant data through innovative technical and community-driven frameworks.**

Privacy preservation and data governance are fundamental to the ethics and scalability of citizen science and crowdsourcing. Projects often collect personal or sensitive information from volunteers, for instance, health metrics or precise geolocations. Thus, robust protocols are needed to protect participant identities, as well as their real-time locations and home addresses, to comply with consent requirements and legal standards. Key research questions include: (1) what security and privacy safeguards are required for citizen science projects that use location-aware mobile phones, cameras, drones, or aerial sensors for data collection to ensure that sensitive imagery or geospatial data are handled in compliance with privacy expectations, laws, and regulations; (2) how might we develop appropriate privacy policies, robust consent mechanisms, and data and AI governance processes[8] throughout a citizen science project's lifecycle; and (3) similarly, how might we develop privacy policies, consent mechanisms, and governance processes in multi-jurisdictional contexts with differing data laws, and data-protection regulations, ensuring projects can operate nationally or globally? Strong governance frameworks also must clarify data ownership, usage rights, and accountability. Innovative models such as community data trusts can provide collective data governance, empowering citizen scientists with greater control over their data, and ensuring it is used for socially beneficial purposes.

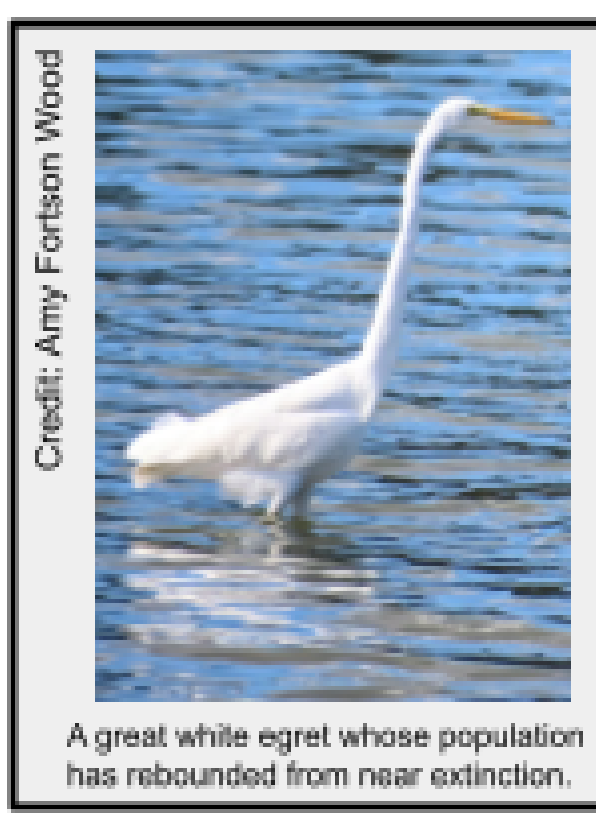

Credit: Amy Fortson Wood

A great white egret whose population has rebounded from near extinction.

**Future Scenario: Protecting Wildlife Data.** In Wyoming, volunteers log wildlife sightings on open platforms like iNaturalist. While public maps engage communities, precise locations risk aiding poachers. Privacy-preserving technologies blur sensitive data while anomaly detection flags suspicious uploads. This balance of openness and protection safeguards endangered species while sustaining public participation.

Developing scalable technical solutions[9] is another priority research area to enable useful data sharing while preserving privacy and confidentiality. Research questions include: (1) how might technical solutions be applied at scale to protect sensitive information without unduly reducing data utility or the accuracy of AI models trained on this crowdsourced data; (2) how might digital twin technology be utilized to enhance security and privacy in citizen science projects, for instance, by creating virtual replicas of sensor networks or data ecosystems to simulate cyberattacks or data leaks; and (3) how might these digital twin simulations help researchers identify potential vulnerabilities, and test mitigation strategies before deploying systems in the real world? This theme is strategically critical: neglecting privacy protections will drive away participants and invite public distrust, whereas effective privacy-preserving practices and transparent governance will enable broader, more inclusive citizen science — including projects involving sensitive data — without compromising individual rights or data integrity.

**4. Data and Model Sovereignty: Balancing local data control with cross-jurisdictional collaboration to ensure fairness and resilience.**

The distributed nature of citizen science and crowdsourcing raises complex issues of data and AI model sovereignty — essentially, who owns, controls, and benefits from the data contributed and the models derived from it? When projects span multiple states, regions, tribes, and/or

[8] e.g., transparent data use policies, granular user consent, and community data stewardship.
[9] e.g., automated anonymization or differential privacy, secure multi-party computation, or homomorphic encryption.

nations, they must navigate restrictions on data sharing and risks of data or models leaking across jurisdictions or to malicious actors. A key research question is how might citizen science research projects ensure data and model sovereignty, giving participants, local communities, and tribal governments greater control over the data they contribute and any AI models derived from it, while still enabling collaborative analysis and innovation on a national or international scale? For example, can emerging approaches like federated learning and decentralized edge computing be harnessed in citizen science research projects to perform data analysis and AI model training locally on participants' devices or within community boundaries? This would keep personal or sensitive data at the source, thereby preserving privacy and data sovereignty, while still enabling collective insights and global collaboration. Also, how might we unlearn or forget learned knowledge from an AI model when data removals are requested? How might we enforce provenance and usage controls so that volunteer-contributed data and resulting models are not misappropriated?

Solutions likely will involve architectures that comply with diverse regulatory requirements and adapt as laws evolve. This theme is strategically important for global citizen science initiatives: without sovereignty safeguards, open data could be exploited by powerful actors for competitive advantage, raising concerns of "AI supremacy" where nations leverage external citizen data for their own gains. Such scenarios can undermine trust and equity. Progress in this area will allow international citizen science collaborations to flourish — sharing insights without raw data exchange — while ensuring communities retain authority over their data and models. In turn, this protects against geopolitical data misuse and fosters equitable, diplomatically sound scientific partnerships.

**5. Openness and Protection: Designing transparent yet secure systems that maintain trust while mitigating risks of exposure.**

Transparency and mutual trust are cornerstones of successful citizen science and crowdsourcing projects. There is a duality of trust in these projects: the scientific community must trust volunteer-collected data and analysis, and participants must trust the scientists leading the project. Research in this area asks how to cultivate and maintain this trust through intentional platform design, protocols, training, and communication. Important questions include: (1) how might we build features that keep volunteers informed throughout and beyond the project lifecycle, and (2) how might we foster mutual trust among all stakeholders? For example, which human-computer interaction (HCI) methodologies and interface designs[10] are most effective at enhancing citizen scientist literacy regarding cybersecurity and privacy risks?

Equally vital is improving transparency of data and AI practices. For example, explainable AI (XAI) and Generative AI tools can be used to improve AI model literacy and annotation training

---

[10] E.g., intuitive UX/UI frameworks, visual analytics tools, interactive training modules

among participants by providing examples of cases that test the boundaries of the AI. These tools also can be used to make any automated analyses more understandable, and potentially to make participants' behavior less vulnerable to malicious manipulation. Furthermore, citizen science projects will need to develop technical standards and data validation algorithms for reliably maintaining the integrity and trustworthiness of data streams from widely deployed, resource-constrained, low-cost sensors, Internet of Things (IoT), including automated anomaly detection, sensor tampering alerts, and validation through cross-sensor correlation methods.

By engaging participants as true partners[11], citizen science initiatives can enhance both trust and data quality. This theme is pivotal because high levels of trust and engagement translate directly into greater future participation and more reliable outcomes. If transparency or accountability is lacking, volunteers may disengage and the public could question the findings; conversely, a culture of openness (when appropriate) and respect strengthens credibility, improves societal uptake of results, and empowers more people to contribute to research.

## Driver 5: Sustaining Infrastructure across Cyber, Data, Human and Physical Layers

The convergence of technology, particularly personal devices and AI, is accelerating the potential of citizen science. At the same time, the distributed, collaborative, long-term, and contextual nature of citizen science also makes it the most demanding use case for 21st-century scientific infrastructure. At its core, infrastructure, not just enthusiasm or expertise, is what enables or constrains citizen science. Whether it is bird counts in a suburban park (*Cornell University*), litter pollution monitoring in the Amazon (Lynch, 2018), or a global network of patients reporting rare symptoms (Schaaf et al., 2024), every phase of public participation[12] in science and engineering relies on systems that must be robust, usable, adaptable, and scalable.

Most citizen science platforms were launched after 2006, coming into existence almost simultaneously with the smartphone. They are built for well-scoped projects and are based on assumptions of the availability of power, network connectivity, and other resources, as well as on access to homogeneous sensors and protocols. Today, citizen science is a key part of how we do science. It routinely involves thousands of people of different backgrounds from various places, sometimes across the country or around the globe. These participants have different skills and reasons for joining. They often collect important data in areas with limited technology. However, our current cyberinfrastructure, network, and data systems are not sufficient to support the full integration of  citizen science projects into scientific workflows (Luettgau et al.,

[11] E.g., through feedback loops, education, and co-design
[12] E.g., data upload, fusion, validation, sharing, interpretation, analysis, visualization, insight inference, and education

2023). Citizen science is truly testing the limits of computer-powered science.

We need to build strong, flexible, and private systems. These systems must work even in places with no internet, few resources, or few sensors. This will help us to meet changing and unique needs. It also requires a mix of computing methods, easy data sharing, and smart, understandable AI tools. The envisioned scientific infrastructure is one that can support both scientific rigor and public participation at large scale as well as high variability across time, space, and individuals. Moreover, the infrastructure needs to enable easy participation, contribution, and use of the tools comprising the infrastructure itself to ensure that it is adaptable to the technology and the endless and ever-changing frontier of science (National Academies of Sciences, Engineering, and Medicine, 2020). The comprehensive scientific infrastructure stack, from hardware and cyberinfrastructure to data and computational frameworks, is an integral part of the future of the scientific process, in a way similar to telescopes, fiber optic networks, and supercomputing centers. A direct investment should be made in research infrastructure to ensure its sustainability and longevity throughout the entire lifecycle of research projects, and the transitions among the projects (OECD, 2025).

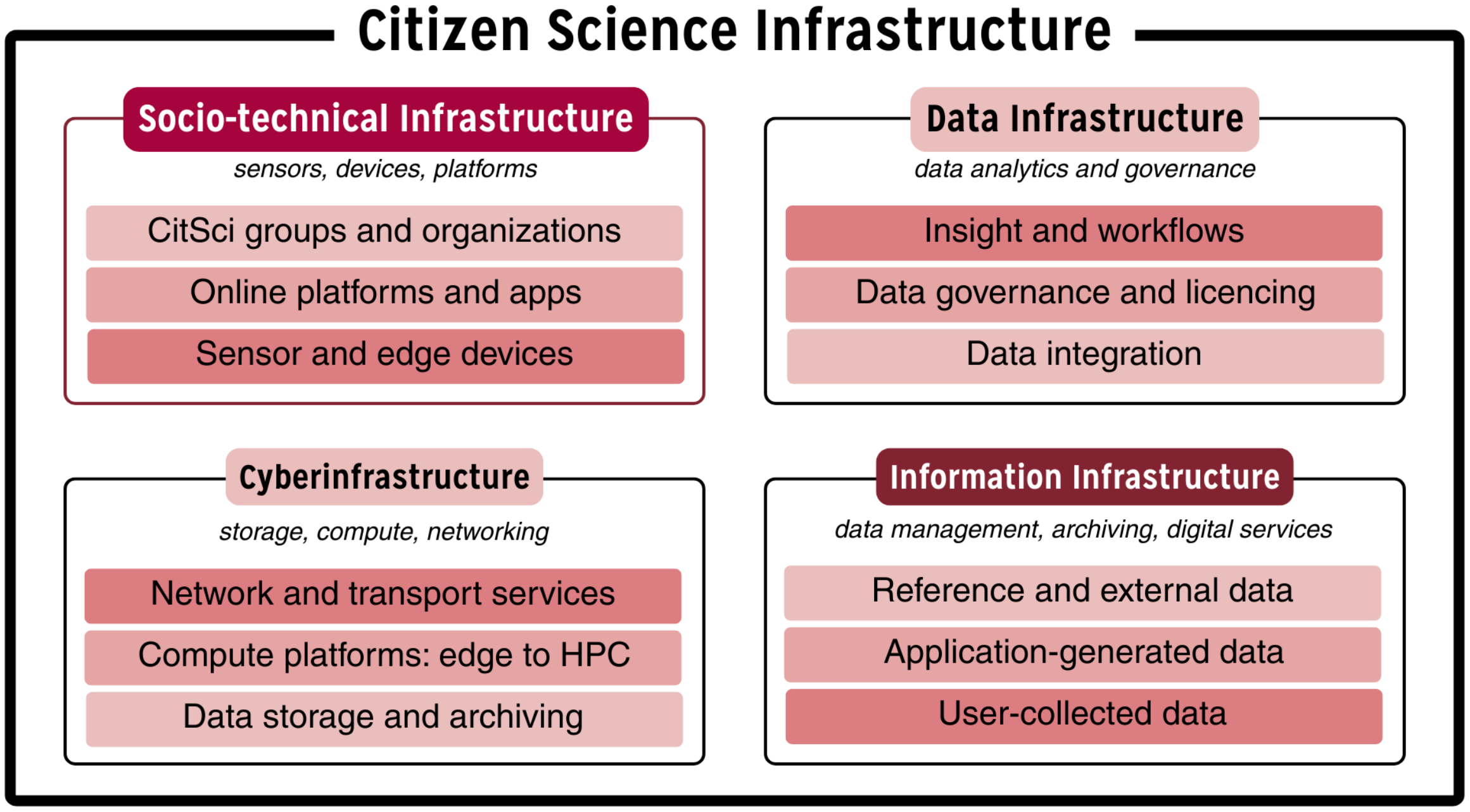

*Figure 2. Science cyber-, information, data, and socio-technical infrastructures, that together comprise the advanced infrastructure for citizen science (Brenton et al., 2018; Yu et al., 2021). The four represented aspects of the infrastructure are distinct but heavily interconnected. Cyberinfrastructure (storage, compute, network) provides the technological foundation; information infrastructure enables the initial data intake and organization while supporting the data infrastructure that provides end-to-end data integration, workflows, and governance; relying on and contributing to each aspect of it is the socio-technical infrastructure of sensor, personal devices, and citizen science platforms.*

Seven key areas for research are identified:

**1. Edge Sensing and Computing: Advancing low-power, distributed technologies to enable data collection and analysis in resource-constrained environments.**

Extreme edge devices (EEDs) are the devices at the furthest endpoints within an edge computing environment (Shi et al., 2016). They typically are positioned at the point of data collection and perform critical computational tasks in environments with limited power, storage, compute, network, and accessibility. Research is needed to advance the design, calibration, and integration of efficient sensors and compute devices that work at the extreme edge in severely resource-constrained environments. In addition, research is needed on the adaptable cyberinfrastructure that spans the continuum from the edge to centralized architectures in an agile and intelligent way (Panda et al., 2024; Iyengar & Pearson, 2024).

**2. Lightweight Machine Learning: Developing compact models that run on personal devices while supporting distributed training and collective insight.**

Modern Machine Learning (ML) is increasingly resource-demanding not just for the training of the models, but also for deployment (Chowdhury et al., 2025; Garg & Kitsara, 2025). In the context of citizen science, ML models often are deployed on personal devices with limited power, storage, bandwidth, and compute resources. There is a growing demand for Machine Learning models that rely on less data and can run on small devices, such as cell phones, pushing advances in efficient model design, compression, distillation, and inference (Wang et al., 2024), and reaching for new model architectures and paradigms. Complementary to that, research also is needed in distributed, shared-resource (e.g., cloud) model training and deployment paradigms to support the distributed and unevenly resourced nature of citizen science.

**3. Hybrid Connectivity: Integrating heterogeneous infrastructures, from local devices to 6G, to ensure reliable, low-latency performance across contexts.**

Citizen science is a distributed endeavor enabled by a variety of local connectivity infrastructures. To support the highly varied connectivity demands with a large range of delay requirements for both upload and download speeds, a new type of intelligent network infrastructure is needed that supports heterogeneous connectivity contexts from the extreme edge to 6G. While not unique to citizen science, the frontier of network research is that which meets the "escalating requirements for ultra-reliable, low-latency communication" (Ishteyaq et al., 2024).

**4. Data infrastructure: Developing AI-ready, interoperable, decentralized data systems so citizen-science contributions can be easily shared and reused.**

Citizen science-enabled research is hindered by the fractured nature of isolated platforms with bespoke data protocols and access processes, and the need for costly post-hoc integration of diverse, unevenly sampled data from fragmented, siloed sources. The majority of citizen science-contributed data remains untapped due to the time and cost of data integration, alignment, engineering, and processing. Moreover, the majority of data is not Findable, Accessible, Interoperable, and Reusable (FAIR) (Wilkinson et al., 2016), which further prevents us from leveraging AI at scale to incorporate citizen science data into scientific workflows. Yet, aggregating data into one source is not the answer. Different data owners and holders are in different jurisdictions with a variety of data licenses and agreements that must be respected (Jimenez, 2020), but could be automatically negotiated (Filipczuk et al., 2022 & Wu et al., 2025). There is a pressing need for an **AI-ready decentralized, coordinated, adaptable, and expandable data infrastructure that advances sustainable, collaborative, open** citizen science and citizen science-enabled scientific research (Nandi et al., 2025).

**Future Scenario: Critical Infrastructure Monitoring.** In Georgia, citizens capture photos, videos, and sensor readings of bridges, water plants, and rail lines. A cloud-based AI system analyzes the data for cracks, corrosion, or vibration anomalies, sending alerts to state engineers. Volunteers extend inspection capacity, while AI ensures scale and accuracy. Together, human input and computational tools keep critical infrastructure safe.

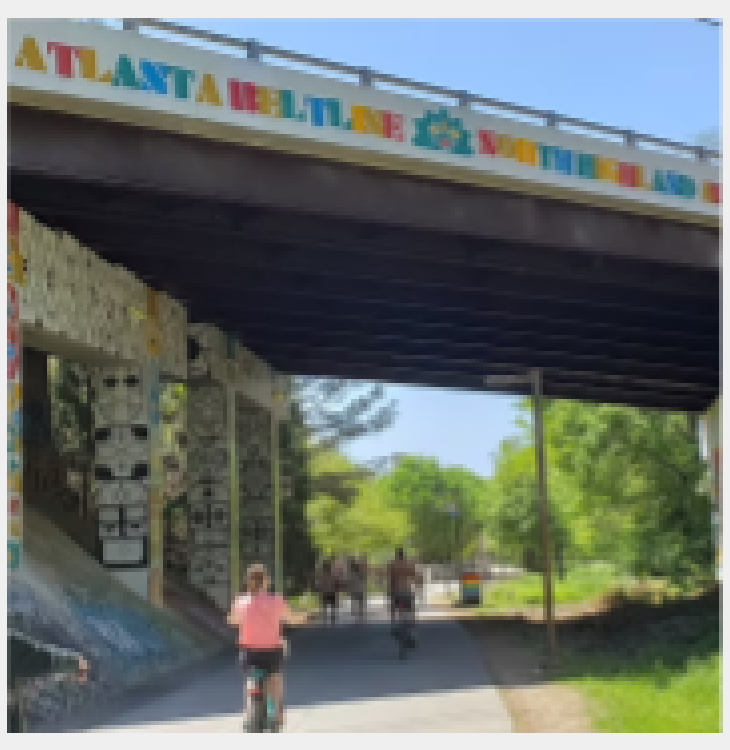

**5. Adaptive Participation: Designing modular, flexible systems that evolve with user needs, and rapidly incorporate new technologies.**

Research is needed to leverage technological advancements to enhance citizen science, ensuring that platforms adapt to the context and user, ensuring inclusive access and participation. Modern advances in adaptive interactive technologies, including LLMs, assistive AI, human-in-the-loop and machine-in-the loop paradigms, and multimodal miniaturized sensors[13] should be explored and leveraged to enable a dynamic and personalized citizen science experience. The coevolution of cyber-, network, data infrastructure (collectively known as the e-infrastructure) and scientific progress needs to respond to the changing nature of citizen science applications and processes (Yu et al., 2021). Research is needed into designing systems that are highly flexible, modular, and easily adaptable to unknown contexts (Bush, 1945) often encountered in citizen science settings. As the technology adoption cycle gets shorter, infrastructure design needs to support the broad/widespread and rapid scaling of new technology and tools (Zhang et al., 2022).

[13] e.g., camera, acoustic, haptic, etc.

**6. Sustainability Models: Establishing robust strategies for long-term maintenance, funding, and improvement of citizen science platforms.**

Digital platforms are often a core component of citizen science projects, and act as both a key tool (or set of tools) used by participants to conduct research as well as acting as connective tissue between agents and stakeholders (Soacha-Godoy et al., 2025). Additionally, citizen science platforms themselves are often infrastructure relied on by researchers to perform their research. However, these key components can be hard to resource. Projects are commonly driven by scientific teams with limited expertise in areas such as software engineering, user interface and user experience design, community management, or instructional design (Liu et al., 2021). Without such skills, teams may not be able to build robust and adaptable infrastructure limiting potential capacity and impact. Additionally, difficulties can arise in managing ongoing support and maintenance, as well as allowing for continuous improvements that take advantage of the latest advancements. Research is needed to develop technical continuity and sustainability models, as well as best practices for development, including potential for cross-sector and cross-stakeholder collaborations.

**7. Practitioner and Participant Training: Building technical and organizational capacity to sustain thriving, technology-enabled citizen science communities.**

Effective citizen science projects are the result of using expertise across multiple disciplines. To support the development of high-quality projects that are able to capitalize on the latest advancements across disciplines and leverage outcomes from prior projects, it is important to establish support for a standalone technical citizen science expertise. Future work should be undertaken to determine effective systems to support practitioners, their learning and skill development, and collaboration.

Participants are the lifeblood of citizen science projects. Special attention is needed for support of participants in the use of cyberinfrastructure, software, digital tools, and platforms throughout the entire project lifecycle, from outreach and recruitment, to onboarding, to skill development, to long-term contribution. Research should be pursued exploring ways to support participants in all stages and accounting for unique challenges across varied groups of participants.

## KEY FINDINGS

The convergence of computational and citizen science research represents a generational opportunity to reimagine how we conduct research, involve the public, and deliver scientific value to society. This CCC workshop surfaced a rich set of ideas for near-term investment and long-term strategic alignment. Rather than treating citizen science and computational science as parallel tracks, the workshop reframed them as interdependent pillars of 21st-century science, each strengthening the other when thoughtfully integrated. Key aspects of this interdependency include:

1. **Mutual reinforcement of AI/ML and citizen science:** A positive feedback loop exists between citizen science and AI/ML in which each can augment and enhance the other, enabling entirely new use cases. Using crowdsourcing and human-in-the-loop methodologies, as well as feedback to models and systems, citizen science benefits AI/ML models by providing trusted, contextualized data across multiple scales. AI/ML benefits citizen science by providing optimized task routing, real-time coaching, and feedback, for example, through judicious deployment of Large Language Models.

2. **Human–computer teaming is essential:** Effective teaming requires balancing machine efficiency and full parameter exploration with human insight and interests, especially in multi-agent, multi-modal, and multi-setting contexts. Humans can catch and explain anomalies, biases, and contextual errors that AI alone misses, interpreting and deciding what is relevant for scientific purposes.

3. **Feedback and interactivity are critical for engagement and data quality:** Personalized, real-time, reciprocal feedback (e.g., via AI tutors, multilingual support, iterative design) is needed for volunteer retention, accuracy, and learning.

4. **Trust is fragile but foundational:** Trust in computational systems hinges on transparency, participatory governance, explainable AI, and clear metrics of mutual respect and accountability between system developers and project participants.

At the same time, we must address key challenges:

5. **Infrastructure as a bottleneck and opportunity:** The needs of citizen science exceed the capacity of current cyberinfrastructure, networks, and data systems more than most scientific domains, exposing the need for next-generation infrastructure (e.g., edge devices, lightweight AI models, decentralized and FAIR data systems, long-term sustainability models). Novel infrastructure could allow for dynamic scaling currently not possible that will accelerate science, provide additional context and new data.

6. **Security, privacy, and adversarial threats are increasing risks:** Citizen science platforms face growing risks from synthetic media, data poisoning, and cyberattacks. Privacy-preserving and sovereignty-respecting governance models are underdeveloped.

7. **Momentum is real but fragmented:** Many successful projects that combine citizen science and computational research exist (e.g., iNaturalist, Mesonet, Stall Catchers, and Zooniverse.org), but lack shared standards, sustainable resourcing (especially for platform support), and coordinated governance — limiting scalability. Further, there is a knowledge-sharing gap between computational researchers and researchers engaged in integrating computational and citizen science, even though they are often investigating similar problems such as in the area of human-centered computing.

## RESEARCH PRIORITIES AND RECOMMENDATIONS TO MEET NATIONAL NEEDS

The convergence of computational science and citizen science represents a generational opportunity to reimagine how we conduct research, involve the public, and deliver scientific value to society. The CCC workshop surfaced a rich set of ideas for near-term investment and long-term strategic alignment. Below, we outline the key research priorities, collaboration pathways, and governance recommendations that emerged from the workshop.

Our five drivers lay out a detailed research agenda to develop a socio-technical ecosystem where citizen participation and computational systems reinforce each other. In summary, citizen science can be transformed when **humans and machines work as complementary partners** (Driver 1), but this requires strong **feedback systems** that keep volunteers engaged and learning (Driver 2). For such collaboration to succeed, it must be built on **trust** — ensuring governance, transparency, and accountability in AI-enabled systems (Driver 3). At the same time, systems must remain **open yet secure**, balancing accessibility with protections for privacy, security, and data sovereignty (Driver 4). Finally, all of this depends on building and maintaining **sustainable infrastructure and long-term community support** — from adaptable project platforms, edge devices, and lightweight AI models to decentralized data systems, connectivity, and long-term community support (Driver 5).

## RECOMMENDATIONS AND ACTIONS

Clear leadership across federal, state, and tribal governments with targeted investments are essential for the United States and its communities — from rural to urban — to fully capitalize on these opportunities and address the challenges. Based on the above findings and the research agenda encapsulated in the five drivers, we articulate a set of recommendations and actions grouped into three key thematic categories.

**National Infrastructure for Convergence**
This theme focuses on the sustained platforms, governance systems, and physical/cyber architecture required to support scalable, trustworthy, and nationwide convergence efforts.

1. **Encourage and incentivize cross-agency collaboration for convergence projects:** Break down silos by integrating citizen science and crowdsourcing with government agency AI, cloud, and technology strategies, and by requiring agencies to co-develop participatory AI projects, maximizing efficiency and impact. (Supported by findings 7 and national mandates)

2. **Create permanent federal funding streams for convergence infrastructure:** Citizen science projects often struggle in the absence of sustainable funding, for example impeding the ability for long-term monitoring projects to support national needs. In addition, funding is scarce for maintenance and evolution of the more centralized platforms that support a multitude of projects. Treat the platforms as national infrastructure (akin to supercomputers) with stable funding for cyber, data, and socio-technical systems; further dedicated funding is needed to ensure individual projects complete analysis, dissemination and evaluation of impacts. (Supported by findings 5, 7)

3. **Build interoperable and AI-ready data infrastructure for participatory sciences:** Create national frameworks and tooling for FAIR (Findable, Accessible, Interoperable, and Reusable) data exchange across citizen science platforms. Emphasize secure APIs, metadata standards, and plug-and-play model training tools for shared use. (Supported by finding 5)

4. **Develop scalable provenance frameworks for participatory AI systems:** Fund research on lightweight, privacy-aware provenance capture that spans human actions, AI model evolution, and distributed execution, enabling reproducibility, auditing, and long-term reuse at the national scale. (Supported by findings 5, 6)

5. **Invest in privacy-preserving and sovereignty-respecting data frameworks:** Implement and evaluate mechanisms to advance federated learning, decentralized data commons, community data trusts, and open standards for interoperability that are core to the design of participatory infrastructure. Formalize the necessary governance and legal structures for secure data use to ensure security, privacy, and local control of sensitive contributions. (Supported by findings 5 and 6)

6. **Establish national guidance for explainable, transparent, and trustworthy AI in citizen science:** Develop research-backed guidelines for interpretable outputs, community consent, and auditability. Require citizen-facing transparency for federally supported participatory AI systems. (Supported by finding 2, 4, 6)

7. **Develop next-generation participatory AI governance:** Co-create frameworks that give participatory science communities real agency over AI deployment decisions, data use, and model evolution — especially for federally funded platforms. This includes mechanisms for consent, auditability, and public oversight. (Supported by findings 4, 6, and 7)

8. **Launch a National Citizen Science & AI Convergence Hub:** Establish a central, virtual hub (funded by the NSF and other agencies) to share tools, standards, best practices, case studies, and training materials, thus reducing duplication and accelerating adoption. Resources could include guidelines for data sharing, model explainability, and responsible use of AI. (Supported by findings 7 and national mandates)

**Core Research for Convergence**
This theme covers the foundational scientific and socio-technical investigations required to advance the field, focusing on developing new models, metrics, and frameworks for human-AI interaction, trust, and accountability.

1. **Encourage and incentivize cross-disciplinary and cross-sectoral collaboration for convergence projects:** Break down silos by investing in strategies that incentivize collaboration and knowledge sharing between computational and citizen science researchers for development of infrastructure and systems of use for both communities, including the use of competition platforms such as Kaggle. Encourage collaboration between academic researchers, federal agencies, industry (especially cloud and AI providers), and local communities to co-design participatory AI platforms. (Supported by finding 7).

2. **Develop Human-AI teaming frameworks for public participation:** Prototype and study new models for complementary collaboration between AI and citizen scientists, particularly incorporating large language models. Specifically investigate novel systems where complementary roles are driven both by multi-agent decisions and community needs. Explore how these systems balance efficiency with human contextual insight in problems such as task assignment, anomaly detection, and data collection or labeling of large datasets. (Supported by findings 1, 2, and 3)

3. **Develop explainable AI for non-expert users:** Research novel user interface and data visualization techniques to make AI decision-making transparent and interpretable to the general public, as well as developers or researchers, enabling the general public to gain fluency with AI concepts and better understand how AI is used. Determine to what extent these techniques build trust and enable broader public engagement with AI-driven platforms. (Supported by findings 1, 2, 3 and 4)

4. **Design real-time feedback systems for citizen science:** Research systems that enable adaptive, multilingual, and just-in-time digital feedback loops that guide users during data collection and analysis. This includes experimentation with LLMs, AR/VR, edge computing, and mobile-first design. Study these systems for improvement in data quality, user engagement, and learning outcomes. (Supported by findings 1, 3, and 3)

5. **Institutionalize evaluation and trust metrics:** Fund research on trust diagnostics, engagement dynamics, and societal benefit indicators to guide iterative improvement and accountability. (Supported by 4, 7)

6. **Advance participatory AI governance models:** Research mechanisms to ensure communities have real governance over how AI is deployed and data are used, including mechanisms for consent, opt-out, accountability, and oversight. Incentivize co-development of toolkits that help projects explain AI behavior to users, including uncertainty visualization, explainable model outputs, and citizen-led model critique workflows. Offer funding (e.g., challenge grants) for co-designed AI tools that are built with community organizations, encouraging broad public participation in scientific research. (Supported by findings 4, 6, and national mandates)

7. **Pilot Human-AI teaming systems across multiple scales and domains:** Fund research as well as deployment testbeds at local and global scales in domains where humans and AI collaborate in real time, including disaster response, health, and environmental monitoring, building scalable models for multi-agent systems. (Supported by findings 1, 2, and national mandates)

**Training and Capacity Building**

This theme focuses on developing the human capital — the skills, knowledge, and organizational structures — needed to create, manage, and participate in convergence projects across all sectors.

1. **Cross-disciplinary training programs:** Develop new academic and practitioner training pathways that combine AI, civic science, cybersecurity, human-centered design, and policy — fostering a new generation of convergence-ready scientists, engineers, and public leaders. (Supported by finding 7 and national mandates)

2. **Leverage citizen science for workforce development:** Integrate computational citizen science into K–12 and continuing education curricula to build STEM, ML/AI skills and civic literacy, aligned with workforce needs in AI and data science. (Supported by findings 3)

3. **Embed real-time feedback systems via AI across citizen science platforms:** Deploy systems incorporating AI that train and guide users, effectively building user capacity and skills while also improving data quality. Leverage LLMs, AR/VR, and mobile-first design for multilingual, adaptive guidance that sustains participation and boosts data quality. (Supported by finding 3)

4. **Broaden participation:** Using AI, bring citizen science into familiar digital and physical environments (apps, games, AR) to expand participation. (Supported by findings 2, 3, and 5)

We also recommend measures to incentivize continued dialogue to capture the momentum from this visioning workshop and build capacity for convergence. This can be accomplished through:

1. **Annual Convergence Research Summits**: Bring together citizen science leaders, computational researchers, public agency representatives, and community advocates to review progress, share outcomes, and set updated priorities.
2. **Dynamic Report Updates:** Publish living documents or digital supplements to this report every 12–18 months, incorporating new research findings, policy shifts, technology developments, and community input.
3. **Online Community of Practice and Knowledge Exchange:** Launch a moderated, cross-sector forum where teams can share use cases, code, infrastructure solutions, and insights in real-time. This will reduce duplication and accelerate innovation.
4. **Evaluation and Metrics Working Group:** Convene a working group to define success metrics for convergence efforts — tracking not only scientific output, but also public trust, equity in participation, and societal benefit.

By strategically converging citizen science and computation, America can usher in a new era of scientific innovation, technological leadership, and public engagement. These coordinated efforts promise not only to sustain, but also significantly amplify the United States' global leadership in science and technology, improve public-sector efficiency, and meaningfully engage millions of citizens in the essential scientific questions of our time, while upskilling them on the use of AI. In the face of global competition, U.S. leadership in these technologies and outcomes is not guaranteed. The time to act decisively, through informed policy and targeted investment, is now.

## CONCLUSION

The **Grand Challenges for the Convergence of Computational and Citizen Science Research Workshop** advanced the national conversation about the future of computing, data sciences, and public engagement in scientific research. Convened by the Computing Community Consortium (CCC) and supported by the National Science Foundation, the workshop brought together an interdisciplinary group of researchers across computing and data sciences, citizen science, and crowdsourcing, as well as adjacent fields. It reflected an urgent and shared recognition: **we are entering a new era where we must investigate how computational systems — particularly AI — and public participation can be combined strategically to address the most complex scientific and societal challenges.**

What made this workshop particularly impactful was its breadth and depth of engagement. Through virtual roundtables, in-person breakout sessions, and collaborative writing, participants surfaced not only the technical opportunities for convergence, but also the educational, infrastructural, and governance considerations. Rather than treating citizen science and computing as parallel tracks, the workshop **reframed them as interdependent pillars of 21st-century science — each strengthening the other when thoughtfully integrated.** This workshop laid the foundation for a national research agenda grounded in both computational research and participatory sciences.

Citizen science is not simply a method of outreach or education. It is a research model that, when integrated with computation, enables scale, inclusivity, and resilience in ways traditional science may not. It can distribute problem-solving across millions of minds and sensors, with real potential for the democratization of data and the ability to foster trust through transparency. Lastly, it can bring science closer to the people it is meant to serve.

The technologies we have developed — such as machine learning, edge computing, cloud infrastructure — are no longer experimental. They are ready to serve society. Their full impact, however, only will be realized when we combine them with the **creativity, curiosity, and local knowledge of the public.**

The vision that emerged from the workshop is bold but achievable: a national research ecosystem where **citizen scientists, computer scientists, data scientists, domain scientists, and computational systems collaborate at scale**, with human and AI teams enhancing one another. This approach reimagines how we monitor our world, respond to disasters, track public health, manage our resources and infrastructure, and even explore space, through **partnerships grounded in mutual trust, shared responsibility, and collaborative intelligence**.

The question is no longer *if* we will converge these fields — but *how quickly and wisely* we choose to do so.

# APPENDICES

- **Appendix A:** The Visioning Process
- **Appendix B:** List of Workshop Participants
- **Appendix C:** Funding Acknowledgements
- Supplementary Data (e.g., survey responses, breakout session notes, visual recordings)
- References and Supporting Documents

## Appendix A: The Visioning Process

The goal of the visioning process was to reveal research areas and opportunities for the convergence of computational and citizen science research. The process unfolded in two phases, first two virtual roundtables to gather input from a broad group, and then a two-day in-person workshop.

Participants were invited directly and via an open invitation posted to numerous mailing lists. Interested participants were asked to submit a 1-page description of their interests, background in citizen science and computational research, and suggestions for important research questions at that intersection. The submissions were used to select invitees for the in-person workshop, and synthesized to create an initial set of topics for discussion.

### *Virtual Roundtable Process*

We started the visioning process with two two-hour virtual roundtable sessions on February 17, 2025, in order to engage participants who were unable to attend the in-person event. One session was held in the morning, and one in the evening (eastern time) to accommodate participation from around the world, bringing together U.S.-based academics and renowned international experts in citizen science. The virtual workshop began with a brief overview of the goals for the sessions led by the organizers. Participants were then divided into breakout groups to explore specific challenges in the field. Each group identified barriers, proposed solutions, and outlined future research questions and directions on the following topic sets:

**Topic Set 1**

- **Computational Citizen Science:** Enhancing citizen science efforts through the use of data analysis, mathematical modeling, computational simulations, and automation of data collection, analysis, and interpretation (e.g., enabling citizen science through AI Assistants).

- **Human-computer Teaming**: Leveraging human and AI collaboration to enhance data analysis and problem-solving (e.g., Real-time Data Processing; Anomaly Detection; Data Quality).

**Topic Set 2**

- **Enabling Citizen Science through Low-cost Sensor Development and Use:** Developing, testing, and evaluating affordable and accessible sensors to empower communities, volunteers, and other contributors.
- **Citizen Science Trust, Equity, Ethics, and Responsible AI:** Exploring ethical and social implications of integrating AI with citizen science (e.g., data bias, quality of AI-derived science, public data for public good, AI-enabled data extraction and incentives).
- **Citizen Science Privacy and Security:** Addressing cybersecurity, security, and mis-/dis-information in citizen science projects and platforms (e.g., masking volunteer information, cognitive security).

**Topic Set 3**

- **Citizen Science Data Cyberinfrastructure, Software, Tools, and the Cloud:** Enhancing citizen science efforts by providing and maintaining robust cyberinfrastructure and cloud computing resources.
- **Cyberlearning through Citizen Science**: Using AI and other computational tools to improve training and education within citizen science and community-engaged science.
- **Increasing Participation at the intersection of Computing and Citizen Science Research:** Reducing participation barriers and fostering broad engagement in citizen science.

Each group rotated through all of the topic sets and provided their input using these questions to frame the discussions:

- What is your vision for the future in the convergence of Computing, Data Sciences, AI and Citizen Science (e.g., open research questions, missing data, or unsolved problems)?
- What are the resources needed to achieve that vision?
- What are examples/use cases/experiences that can illuminate that vision or highlight a needed resource?

We used the online tool Mural to collaborate in real time with the participants. An example of a completed Mural board is shown below.

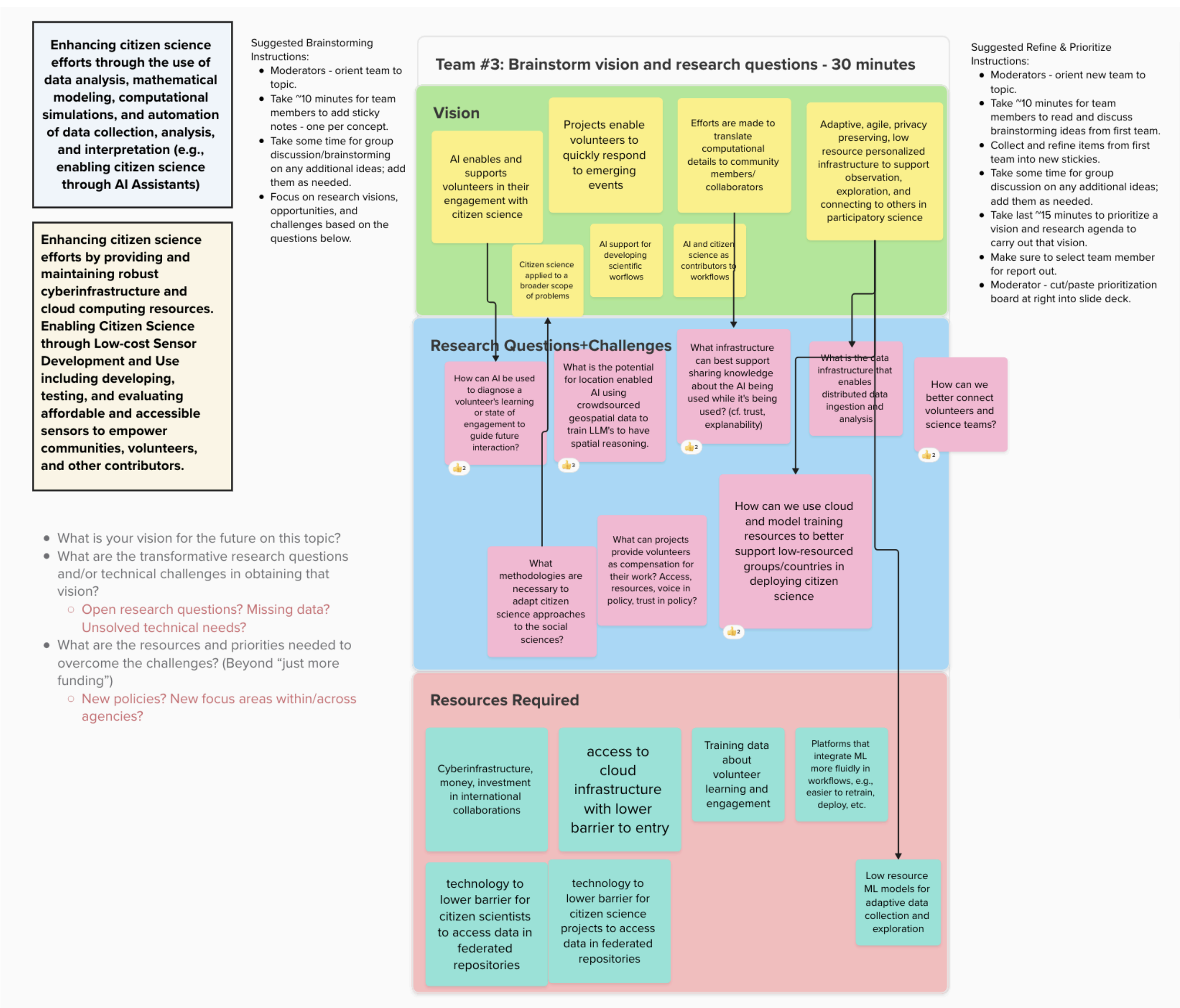

After the sessions concluded, the notes from the discussions were summarized for inclusion in the final report. The summary yielded the following seven overarching themes.

- **AI as a Partner, Not a Replacement.** In citizen science, AI should *augment* volunteer contributions — not overshadow them. Tools must be explainable and designed to foster meaningful engagement, allowing participants to understand and shape project outcomes.
- **Access & Equity.** Inclusive citizen science requires mobile-accessible platforms, low-resource AI models, and infrastructure aligned with FAIR and CARE principles. This ensures equitable participation across communities, especially those historically underrepresented in science.
- **Learning & Capacity Building.** AI can transform citizen science into a space for learning by providing tutoring, real-time feedback, and personalized skill development.

Addressing barriers like language and digital literacy is essential for broader participation.

- **Ethics & Trust.** Citizen science must prioritize ethical AI use — protecting privacy, respecting cultural contexts (including Indigenous data sovereignty), and embedding transparency into data collection and analysis workflows.
- **AI-Human Collaboration.** Next-generation citizen science platforms should enable fluid collaboration between humans and AI, using adaptive tools that support creativity, contextual judgment, and human oversight at every step.
- **Partnerships & Policy.** Scalable, responsible citizen science efforts require cross-sector partnerships, updated ethical frameworks, and alignment with global goals like the UN SDGs to ensure lasting impact.
- **Vision Forward.** Participants envisioned a future where AI empowers citizen scientists — enhancing inclusion, education, and real-time collaboration to build a more ethical, equitable, and impactful citizen science ecosystem.

### *In-person Workshop Process*

The CCC Grand Challenges for the Convergence of Computational and Citizen Science Research Workshop was held in-person in Washington DC on April 7-9, 2025. The full workshop agenda is included below. The workshop started on the afternoon of the first day with introductions and a panel on the state of the field and emerging research questions at the intersection of Citizen Science and Computing as a level-setting activity. The morning of the second day included two talks on the state of the field, one addressing computational research with respect to citizen science and second, citizen science research with respect to computation, to provide all attendees with an appreciation of the topics.

The majority of the workshop time was occupied by a series of four breakout sessions for small group discussion of various topics. The four breakout sessions all had 6 groups of approximately 7 participants and 1 workshop organizer/moderator. Interactive writing tools enabled real-time collaboration, and facilitators guided brainstorming and synthesis. At the end of each breakout, a representative from each group shared key takeaways in a full-group discussion, fostering cross-cutting insights and actionable next steps.

Below are more details about each session.

- Breakout I: What will the next revolution at the nexus of computational and citizen science/crowdsourcing look like?
    - Group curation: An interdisciplinary range of experts was assigned to 6 topic sets at the intersection of computing and citizen science. The topics were derived from the topics discussed in the virtual sessions. The goal of the first breakout was to encourage broad and divergent thinking, hence the assignment of discussants from diverse research areas.

- Breakout Table 1: Computational Citizen Science
- Breakout Table 2: Human-Computer Teaming
- Breakout Table 3: Citizen Science Computing Trust, Equity, Ethics, and Responsible AI
- Breakout Table 4: Citizen Science Computing/AI, Privacy, and Security
- Breakout Table 5: Cyberinfrastructure, Software, Data, and Sensors for Citizen Science
- Breakout Table 6: Citizen Science through Cyberlearning

- Output: Each group developed no more than 6 research questions related to the topic.
- During the report out, flipcharts were provided for each topic so that during the break between sessions, members of other groups could add ideas on sticky notes to other group's boards to encourage cross-pollination.

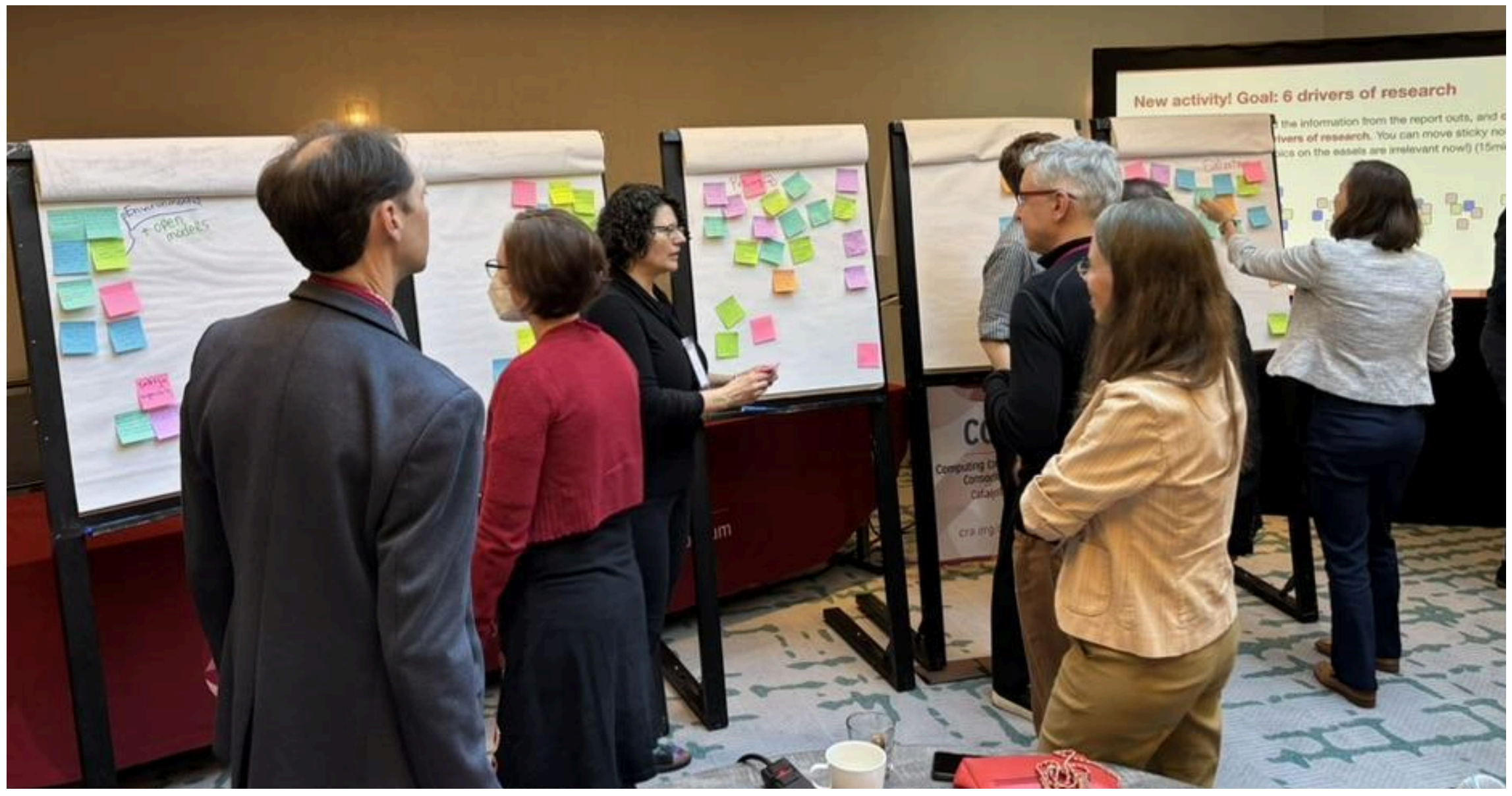

- Breakout II: Convergence of computational and citizen science/crowdsourcing research
  - Group curation: Intra-disciplinary experts were assigned to the same 6 topic sets.
    - While the group members changed for breakout II, the topic and moderators stayed the same, so the ideas from the previous group, plus the additional ideas generated from other participants during the report out time, were preserved and considered for breakout II. The goal of the second breakout was to narrow in on a set of important and feasible questions, hence the assignment of discussants in a focused research area.
  - Output: In new groups, additional research questions were brainstormed, and then the combined list was narrowed down to 5 research questions.

- During the report out, participants were again encouraged to add ideas to the flipcharts and synthesize by grouping related ideas together on the boards.

- After the first full day of discussions, the workshop organizers transcribed all of the notes from the boards. From the generated ideas, they synthesized 6 “research drivers” for further discussion.
  - Driver 0: What makes citizen science unique, and how is it evolving in the age of AI? (Presented in the introduction to the report)
  - Driver 1: How do we enable extreme heterogeneity in multi-agent, multi-modal, multi-dimensional, multi-participant, multi-environment systems (including all combinations of human/machine/digital/physical combinations)?
  - Driver 2: In what ways can feedback mechanisms like interactivity, agency, and just-in-time delivery be optimally implemented?
  - Driver 3: How can multi-partite trust be established and maintained between citizen scientists, scientists, decision-makers, and computational systems (considering safe AI, governance, security, scientific validity of inferences drawn from citizen science input, system evaluation, bias, missing data, user experience)?
  - Driver 4: What are the key challenges and potential solutions for ensuring privacy and security in open, participatory systems involving citizen science?
  - Driver 5: How can we build, deliver and sustain adaptive computational infrastructure (cyber, data, sensors, software, human, etc.)?

- Breakout III (start of day 3): Fundamental Citizen Science & Computing opportunities at the nexus of computational and citizen science/crowdsourcing (Applications | Synthesis and prioritization of Breakout II research drivers)
  - Group curation: Participants voted on which drivers they wanted to spend time on for Breakout III. Based on the results, the workshop organizers assigned participants to groups, ensuring distribution across “drivers.”
  - Output: answer the following questions for the assigned “research driver”:
    - What is missing in this driver?
    - Lay out the path that we need to drive on?
    - If we answered this question, what would it help us achieve in citizen science?
    - Who and what resources are needed?
    - What examples demonstrate that the approach/solution works well, and how it might be applied to citizen science?

- Breakout IV: Report Writing: Synthesis and Recommendations
  - Group curation: while they were given the opportunity to change “drivers” if they wanted to, most workshop participants stayed in their same groups as Breakout III in order to continue fleshing out content on the topic.
  - Output: draft text for the driver’s section of the workshop report.

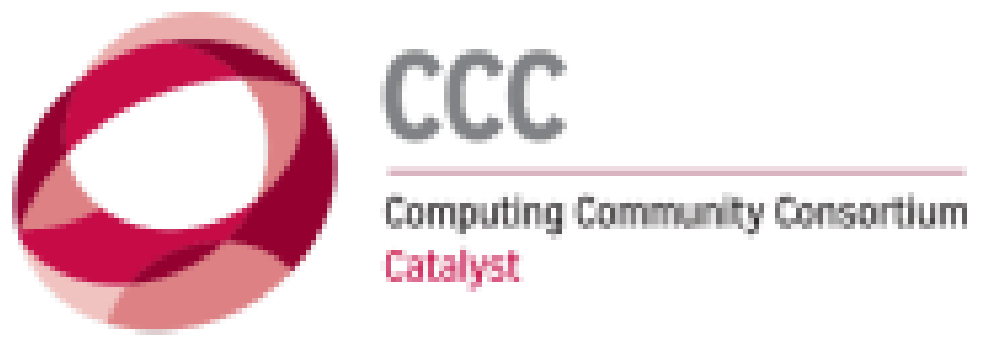

**Grand Challenges for the Convergence of Computational and Citizen Science Research** CCC Workshop DRAFT Agenda

*Viceroy Hotel, 1430 Rhode Island Ave NW, Washington, DC 20005*

***Monday, April 7***

*All times in EST*

| | |
|---|---|
| 4:00 PM | COFFEE/SNACKS |
| 4:10 PM | Introduction |
| 4:30 PM | Participant Introductions (lightning slides) |
| 5:30 PM | Introductory Activities and Material |
| 6:30 PM | Conclude Day 1 |
| 7:00 PM | Reception with Heavy Hors d'oeuvres + Meet & Greet \| (organized by Workshop) |

***Tuesday, April 8***

| | |
|---|---|
| 7:30 AM | BREAKFAST |
| 8:30 AM | Welcome to Day 2 |
| 8:45 AM | State of the Field |
| 9:45 AM | BREAK |
| 10:15 AM | Breakout I: What will the next revolution at the nexus of computational and citizen science/crowdsourcing look like? |
| 12:00 PM | Report Out and Group Discussion I |
| 1:00 PM | LUNCH \| (in restaurant across the hall) |
| 2:00 PM | Breakout II: Applications at the convergence of computational and citizen science/crowdsourcing research. |
| 3:45 PM | BREAK |
| 4:15PM | Report Out and Group Discussion II |
| 6:00 PM | Conclude Day 2 |
| 6:30 PM | DINNER \| (organized by Workshop) |

***Wednesday, April 9***

| | |
|---|---|
| 7:30 AM | BREAKFAST |
| 8:30 AM | Review Day 2 / Logistics Day 3 |
| 9:00 AM | Breakout III: Fundamental Citizen Science & Computing opportunities at the nexus of computational and citizen science/crowdsourcing \| Opportunities for Government and Philanthropy to maintain leadership |
| 10:30 AM | Report Outs and Group Discussion III |
| 11:00 AM | BREAK |
| 11:30 AM | Breakout IV- Report Writing: Synthesis and Recommendations |
| 1:00 PM | WORKING LUNCH |
| 2:00 PM | Wrap Up and Closing Remarks |
| 2:30 PM | Adjourn |

## Appendix B: List of Workshop Participants

| First Name | Last Name | Affiliation |
|---|---|---|
| Nitin | Agarwal | COSMOS Research Center, University of Arkansas at Little Rock |
| Sara | Beery | MIT |
| Tanya | Berger-Wolf | Ohio State University |
| Nataly | Buslón | Pompeu Fabra University |
| Tracy | Camp | Computing Research Association |
| Ryan | Carney | University of South Florida |
| Varun | Chandola | National Science Foundation |
| Seth | Cooper | Northeastern University |
| Kevin | Crowston | Syracuse University |
| Yi | Ding | University of Texas at Dallas |
| Lucy | Fortson | University of Minnesota/Zooniverse |
| Jen | Frazier | The Moore Foundation |
| Kobi | Gal | Ben-Gurion University |
| Sanjana | Gautam | University of Texas at Austin |
| Alina | Gerall | Computing Research Association |
| Catherine | Gill | Computing Research Association |
| Josh | Greenberg | Alfred P. Sloan Foundation |
| Haley | Griffin | Computing Research |

| | | Association |
|---|---|---|
| Muki | Haklay | University College London |
| Peter | Harsha | Computing Research Association |
| Haym | Hirsh | Cornell University |
| Carolynne | Hultquist | University of Canterbury |
| Corey | Jackson | University of Wisconsin |
| Aggelos | Katsaggelos | Northwestern University |
| Harmanpreet | Kaur | University of Minnesota - Twin Cities |
| Kevin | Kells | University of Ottawa |
| Sola | Kim | Arizona State University |
| Elizaveta (Lee) | Kravchenko | Northeastern University |
| Marc | Kuchner | NASA |
| Thai | Le | Indiana University |
| Matt | Lease | University of Texas at Austin |
| Daniel | Lee | Schmidt Sciences |
| Chris | Lintott | University of Oxford |
| Rosalia | Maglietta | Italian National Research Council |
| Mary Lou | Maher | Computing Research Association |
| Pietro | Michelucci | Human Computation Institute |
| Joy | Ming | Cornell University |
| Beth | Mynatt | Khoury College at Northeastern University |
| Peder | Nelson | Oregon State University |

| Dara | Norman | NOIRLab |
|---|---|---|
| Julia | Parrish | University of Washington |
| Daniel | Pittman | Metropolitan State University of Denver |
| Janet | Prevey | US Geological Survey |
| Jonathan | Romano | Frameshifter |
| Saiph | Savage | Northeastern University |
| Karen (Kat) | Schrier | Marist University |
| Carrie | Seltzer | iNaturalist |
| Kumba | Sennaar | Science Philanthropy Alliance |
| Lea | Shanley | International Computer Science Institute at Berkeley |
| Laura | Trouille | Adler Planetarium - Zooniverse |

## Virtual Roundtable Participants:

Segun Adebayo, Bowen University
Kevin G Crowston, Syracuse University
Hugh Dickinson, The Open University
Heather A. Fischer, Oregon State University
Chelle Gentemann
Carolynne Hultquist, University of Canterbury
Harmanpreet Kaur, University of Minnesota - Twin Cities
Sarah Kirn, NASA Citizen Science
Laure Kloetzer, University of Neuchâtel, Switzerland
Matt Lease, University of Texas at Austin
Yiftach Nagar, Academic College of Tel Aviv-Yaffo & University of Haifa
Saiph Savage, Northeastern University
Sven Schade, European Commission, Joint Research Centre (JRC)
Daniel Sifael, University of Dodoma
Alex Szalay, Johns Hopkins University

Michela Taufer, University of Tennessee-Knoxville

### *Participant Demographics*

1. Total number in both virtual sessions (not including CRA staff, but including 6 organizers): **24 (roundtable 1) + 12 (roundtable 2) = 34 total**

2. Total number for in-person (not including CRA staff, but including organizers): **46**

3. Disciplinary focus — computation or citizen science (or clearly both) — for the in-person workshop, comprised of: 13 computer science, 13 citizen science, 20 with some combination of background knowledge in computer science and citizen science.

**The table below summarizes:**

4. For both virtual and in person sessions, the break down by institution type (academia, NGO, federal agency, foundation, industry)

5. Country of origin

| | Academic | NGO (Museums, Societies, etc.) | Federal Agency | Foundation | Industry | Total |
|---|---|---|---|---|---|---|
| **Virtual*** | 20 | 4 | 1 | 0 | 1 | 26 |
| **In Person**** | 32 | 5 | 5 | 3 | 1 | 46 |

*majority US, with 9 representatives from Africa, Israel, Europe, New Zealand, and the UK
**majority US, with 7 representatives from Canada, Israel, New Zealand, and the UK

## Appendix C: Funding Acknowledgements

United States National Science Foundation

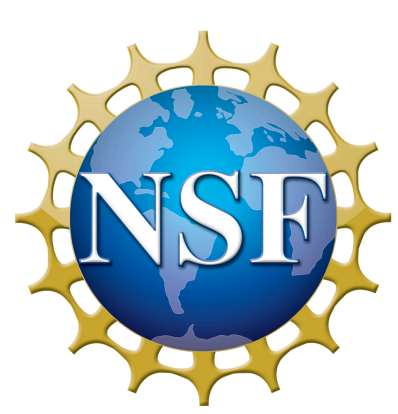

The workshop was supported by the Computing Community Consortium through the U.S. National Science Foundation under Grant No. 2300842. Any opinions, findings, and conclusions or recommendations expressed in this material are those of the authors and do not necessarily reflect the views of the U.S. National Science Foundation.